\documentclass[12pt,a4paper]{article}
\usepackage[vcentermath,enableskew]{youngtab}
\usepackage{amsfonts}
\usepackage{amsmath}
\usepackage{bm}
\usepackage{amssymb}
\usepackage{fullpage}
\def\Dots{\cdot\cdot}
\newcommand{\unit}{\ID}
\newcommand{\boplus}{\bigoplus}

\newcommand{\ID}{\mathbf{1}}

\newcommand{\CP}{\mathbb{C}\mathbb{P}}
\newcommand{\Gr}{\mathbb{G}\rm{r}}
\newcommand{\tA}{\widetilde A}

\newcommand{\Ssec}{section\ }

\newcommand{\ket}[1]{\left| \! \right. #1\left. \! \right\rangle}
\newcommand{\bra}[1]{\left\langle \! \right. #1\left.\! \right| }

\newcommand{\eq}{\begin{equation}}
\newcommand{\qe}{\end{equation}}
\newcommand{\eqa}{\begin{eqnarray}}
\newcommand{\qea}{\end{eqnarray}}
\newcommand{\al}{\alpha}
\newcommand{\bt}{\beta}
\newcommand{\g}{\gamma}
\newcommand{\dl}{\delta}
\newcommand{\ta}{\theta}
\newcommand{\e}{\epsilon}
\newcommand{\avec}{\ensuremath{\bm{\alpha}}}
\newcommand{\bvec}{\ensuremath{\bm{\beta}}}
\newcommand{\svec}{\ensuremath{\bm{\sigma}}}
\newcommand{\tauvec}{\ensuremath{\bm{\tau}}}

\newcommand{\gvec}{\ensuremath{\bm{\gamma}}}

\newcommand{\vv}[1]{\ensuremath{\bm{#1}}}
\newcommand{\dd}[2]{\frac{\partial #1}{\partial #2}}
\newcommand{\non}{\nonumber}

\newcommand{\comment}[1]{}
\begin{document}

\title{\vspace*{-0.1cm}\hfill \raise 20pt \hbox{\small DIAS-STP-06-18}\\
\date{November 19, 2006}
Noncommutative vector bundles over fuzzy $\CP^N$ and their covariant derivatives}
\author{Brian P. Dolan,$^{1,2}$\footnote{email: bdolan@thphys.nuim.ie}~
Idrish Huet,$^{1,3}$\footnote{email: mazapan@stp.dias.ie}\\
Se\'an Murray $^{1,2}$\footnote{email: smury@stp.dias.ie }\
\; {and}\;
Denjoe O'Connor$^1$\footnote{email: denjoe@stp.dias.ie}
\\
\\
$^{\rm 1}$ \it \normalsize School of Theoretical Physics,\\
\it \normalsize Dublin Institute for Advanced Studies,\\
\it \normalsize 10 Burlington Road, Dublin 4, Ireland.\\
\\
{\it \normalsize $^{\rm 2}$ Department of Mathematical Physics}, \\
{\it \normalsize NUI Maynooth, Maynooth, Co. Kildare, Ireland.}\\
\\
{\it \normalsize $^{\rm 3}$ Depto de F{\'\i}sica}, \\
{\it \normalsize Centro de Investigaci\'on y de Estudios Avanzados del IPN}, \\
{\it \normalsize Apdo. Postal 14-740, 07000 M\'exico D.F., M\'exico.}
\\
\\
}
\maketitle
\begin{abstract}
We generalise the construction of fuzzy $\CP^N$ in a manner that allows us
to access all noncommutative equivariant complex vector bundles over this space.
We give a simplified construction of polarization tensors on $S^2$ that
generalizes to complex projective space, identify Laplacians and natural noncommutative
covariant derivative operators that map between the modules that describe
noncommutative sections.  In the process we find a natural generalization
of the Schwinger-Jordan construction
to $su(n)$ and identify composite oscillators that obey a Heisenberg
algebra on an appropriate Fock space.
\end{abstract}
\vfill\eject
\section{Introduction}
\label{Introduction}
Noncommutative geometry \cite{Connes:1994yd,GraciaBondia:2001tr,
Madore:2000aq,Landi:1997sh} has become an active area of research in
recent years providing potential new physics due to
modifications of the underlying structure of spacetime
\cite{Balachandran:2005eb,Doplicher:1994tu}. Such geometries naturally
arise in string theory \cite{Seiberg:1999vs}. A closely related
development is that of fuzzy physics where finite matrix algebras are
used to approximate the algebra of functions on a manifold. Action
functionals built from these matrix algebras provide an alternative to
lattice actions in the regularization of field theories and are
especially natural for field theories on noncommutative spaces. Fuzzy
spaces have also been pursued as potential spaces for the internal
space in Kaluza-Klein reductions
\cite{Aschieri:2003vy,Dolan:2002af,Dolan:2002ck}.  See
\cite{Balachandran:2005ew} for a review of aspects of the fuzzy
approach and \cite{Douglas:2001ba,Szabo:2001kg} for noncommutative field theory.
The connection between matrix models and string theory is reviewed in
\cite{Taylor:2001vb} and the relation of fuzzy geometry to the
quantum Hall effect is reviewed in \cite{Karabali:2006eg}.

As a regularization of quantum field theory, the fuzzy approach replaces
the infinite number of degrees of
freedom of the Euclidean quantum field with the finite
number of degrees of freedom associated with a direct sum of modules
over a finite simple matrix algebra, where the finite matrix algebra
approximates the algebra of functions of the underlying compact
Euclidean space.  Of course, by Wedderburn's theorem all simple finite
dimensional matrix algebras over the complex numbers are isomorphic
and characterized only by the matrix dimension, so the geometry must
be encoded directly in the action of the field theory. For
scalar fields, it is the Laplacian that does this and hence, once one
has specified the allowed matrix dimensions and the action for a free
scalar field, the fuzzy, and hence limiting commutative, geometries are
determined. In the case of spinor fields, the relevant operator is the Dirac
operator which encodes the spin geometry \cite{Connes:1994yd} of the space.

A simple method of constructing a large variety of fuzzy spaces is as
coadjoint orbits of compact Lie groups. In this way, the fuzzy space
preserves all the isometries of the limiting commutative geometry even
at the finite level and the commutative geometry
is recovered in the limit of infinite size matrices\footnote{Due to the
presence of all the continuous isometries the fuzzy theory is already
continuum.}.  This
is a radical shift from the canonical lattice regularisation that is
already a highly developed tool for quantum field theory. Its
potential advantages are, however, sufficiently promising that it
warrants attention in its own right. The principal advantages arise
for models with Fermions, where the Dirac operators are significantly
simpler than their lattice relatives and furthermore the fuzzy models
avoid the difficulties of Fermion doubling
\cite{Balachandran:1999qu}. Preliminary numerical studies of
scalar field theories on fuzzy spaces have been performed in
\cite{Martin:2004un,GarciaFlores:2005xc,Medina:2005su,Panero:2006bx,Panero:2006cs}.
Fuzzy gauge fields on $S^2_F$ have been discussed by several authors
\cite{Grosse:2001ss,Steinacker:2003sd,Castro-Villarreal:2004vh} and on
$\CP^2$ in \cite{Grosse:2004wm}. Recent
numerical studies  of such gauge theories can be found in
\cite{O'Connor:2006wv} and \cite{Bietenholz:2006cz}.

The archetypical fuzzy space is the fuzzy sphere
\cite{Berezin:1974du,Hoppe:1982,Madore:1991bw},
though any manifold which can be generated as the coadjoint orbit of
a compact Lie group should have a fuzzy description. See Arnlind {\it
et al} \cite{Arnlind:2006ux} for a fuzzy torus construction not based
on coadjoint orbits.\footnote{In this torus example the relevant
Laplacian has yet to be constructed.}  Explicit descriptions of fuzzy
$\CP^N$, fuzzy unitary Grassmannians and fuzzy complex quadrics
already exist \cite{Balachandran:2001dd,Dolan:2001mi,Dolan:2003th}.
There are also constructions
of ``fuzzy'' spheres of dimension greater than two
\cite{Ramgoolam:2001zx,Dolan:2003kq}
but all of these models involve additional degrees of freedom and
require the introduction of some technique to decouple these modes.

To pin down the geometry, one must focus on the scalar Laplacian
(or Dirac operator).
In the literature there exist two distinct prescriptions for the
Laplacian on the fuzzy sphere. The simplest prescription takes the
Laplacian as the quadratic Casimir $\Delta=\hat{\cal L}_a\hat{\cal
L}_a$ where $\hat{\cal L}_a=[\hat L_a,\cdot]$ and the $\hat L_a$ are
the generators of $su(2)$ in the irreducible representation of
dimension $L+1$ where $L$ is any positive integer.  This Laplacian has
a natural generalization to $\CP^N$, where again the quadratic Casimir
of $SU(N+1)$ in the irreducible representation corresponding to the
matrix size is used. However, in \cite{Grosse:1995jt} an alternative
prescription for the Laplacian is given in the case of $S^2_F$. Here
the Laplacian is given as $\Delta=\frac{1}{2}(
\hat K_{-}\hat K_{+}+\hat K_{+}\hat K_{-})$ in terms of operators (see
(\ref{su2KsDefined}) below) $\hat K_{-}$ and $\hat K_{+}$ which
together with $\hat K_0$ satisfy the $su(2)$ algebra. In the
commutative case differential operators corresponding to $\hat {\cal
L}_{a}$ are the right invariant generators of $SU(2)$ on itself,
while those corresponding to $\hat K_{+},\hat K_{-}$ and $\hat K_0$
are the corresponding generators of the left invariant vector fields.
On coset spaces $G/H$ the left and right invariant generators of $G$ play very
different roles: The right invariant generators act as Lie derivatives while the
left invariant generators become covariant derivatives.  In the fuzzy setting
this is reflected in the fact that the operator images of right invariant generators
{\it i.e.}\ $\hat {\cal L}_a$, provide the
generators of the adjoint action of the group $G$ acting on the matrix
algebra while the operators $\hat K_{+}$ and $\hat K_{-}$ change the matrix
sizes with $\hat K_{+}$ mapping from square $(L+1)\times(L+1)$ matrices
to non-square $(L+2)\times L$ matrices and $\hat K_{-}$
mapping from $L\times(L+2)$ matrices.  These latter
matrices can be interpreted as projective modules over the fuzzy
sphere and allow one to access line bundles over the sphere in a very
natural way \cite{Grosse:1995jt}.  Also, it is straightforward to
construct the Dirac operator and action functionals for spinors once
these operators are known. The principal goal of this paper will be to
give the construction of the corresponding operators for any complex
projective space.  We will see that this will involve us in an
alternative construction of $\CP^{N}_F$, which opens up a variety
of possibilities for the addition of structure to these spaces.

In section {\bf 2} we review the construction of both $S^2_F$, and of
topologically nontrivial field configurations on this space and
review the construction of the operators $\hat K_\pm$. Along the way
we give a novel construction of polarization tensors for both square
and non-square matrices (see \cite{Presnajder:1999ky} for the
standard construction).
In section {\bf 3} we repeat the construction for $\CP^N$ ending the section with
a brief description of noncommutative line bundles over $\CP^N$.
Section {\bf 4} introduces composite operators which surprisingly
end up obeying the Heisenberg algebra on an appropriate reduced Fock space.
The setting involves a natural generalization of the Schwinger-Jordan construction
to $su(n)$.
Section {\bf 5} gives the operators $\hat K_\imath$ and $\hat K_{\bar \imath}$ that
generalize $\hat K_\pm$ and map between noncommutative vector bundles and describes the
modules corresponding to these bundles.
Section {\bf 6} contains our conclusions. Some technical
results needed in the text are obtained in appendices.

\section{The Fuzzy Sphere, $S^2_F$ and its noncommutative line bundles}
\label{CP1-Fuzzy}

We begin by focusing on the fuzzy sphere $S^2_F\cong \CP^1_F$ using an approach
which easily generalizes to other spaces. The generalization to $\CP^N$ will be
pursued in subsequent sections.

Let $a^{\alpha}$, $\alpha=1,2$ be a doublet of annihilation operators
that annihilate the Fock vacuum $|0\rangle$
and let $a^\dagger_\beta$ (the Hermitian conjugate of $a^\beta$)
be a conjugate pair of creation operators with the two doublets satisfying
the Heisenberg commutation relations\footnote{It will be convenient to use the
metric $\delta_{\alpha{\overline
\beta}}$ and its inverse $\delta^{\alpha{\overline \beta}}$ on $\mathbb{C}^2$
to raise and lower indices, so that
$a^\alpha =\delta^{\alpha\overline{\beta}}a_{\overline{\beta}}$.
Of course the distinction between upper and
lower case indices for $su(2)$ is somewhat trivial as the
representations are unitarily equivalent. However, for later use it
will be convenient to maintain a consistent notation that extends to
$su(N+1)$.}
\eq
[a^{\alpha},a^{\beta}]=[a^\dagger_{\alpha},a^\dagger_\beta]=0
\qquad{\rm and}\qquad [a^{\alpha},a^\dagger_\beta]=\delta^\alpha_\beta\ .
\qe
The Fock space ${\cal F}$ freely generated by the creation operators,
$a^\dagger_{\alpha}$, is spanned by the orthonormal vectors
\eq
|n_1,n_2\rangle=\frac{1}{\sqrt{n_1!n_2!}}{(a_1^\dagger)}^{n_1}{(a_2^\dagger)}^{n_2}|0\rangle .
\qe
The Schwinger--Jordan construction then gives operators
\eq
\label{u2SchwingerJordanops}
\hat N = a^\dagger a \qquad {\rm and}\qquad \hat L_a=a^\dagger\frac{\sigma_{a}}{2}a
\qe
which satisfy the $u(2)$ algebra
\eq
\label{u2SchwingerJordanalg}
[\hat L_a,\hat L_b]=i\epsilon_{abc}\hat L_c, \qquad [\hat L_a,\hat N]=0.
\qe
The raising and lowering
operators $\hat L_{\pm}=\hat L_1\pm i \hat L_2$ and $\hat L_0=\hat L_3$ are explicitly
\eq
\hat L_+=a_1^\dagger a^2, \qquad \hat L_-=a_2^\dagger a^1,
\qquad \hat L_0=\frac{1}{2}(a_1^\dagger a^1-a_2^\dagger a^2),
\qe
and the algebra can equally be written
\eq
[\hat L_0,\hat L_{\pm}]=\pm L_{\pm},\quad [\hat L_{+},\hat L_{-}]=2\hat L_0,
\quad [\hat N,\hat L_{\pm}]=[\hat N,\hat L_0]=0.
\label{su2_in_raising_lowering_basis}
\qe
Since the $\hat L_a$ commute with $\hat N$ we can decompose ${\cal F}$ into  a direct
sum of eigenspaces of $\hat N$ as
\eq
 {\cal F}={\mbox{\scriptsize$\displaystyle{\boplus^{\infty}_{L=0}}$}}{\cal F}_L.
\qe
The subspace of states ${\cal F}_L$
of the Fock space ${\cal F}$ is the span of the $L+1$ vectors
$|n_1,n_2\rangle$ with $n_1+n_2=L$, {\it i.e.}\
\eq
{\cal F}_L=span\{|n_1,n_2\rangle\ | \quad n_1+n_2=L\}.
\qe
This space is a representation space of the unitary irreducible $L+1$ dimensional
representation ($spin-\frac{L}{2}$) of $su(2)$ on which the
number operator $\hat N$ and quadratic Casimir $\hat L_a\hat L_a$ take the numerical values
$L$ and $\frac{L}{2}(\frac{L}{2}+1)$,  respectively.

Equally one can consider the dual Fock space ${\cal F}^*$ with vacuum
vector ${\langle0|}$ where $\langle 0|0\rangle=1$ and the restricted dual subspaces
${\cal F}_L^*$. The algebra associated with the fuzzy geometry is realized as the linear span of
${\cal F}_L\otimes{\cal F}_L^*$; it is isomorphic to the algebra of
$(L+1)-{\rm dimensional}$ matrices.  The norm on the matrix algebra is
taken to be the trace norm. Define $\hat {\cal L}_a=\hat L_a^{\rm L}-\hat L_a^{\rm R}$,
with $\hat L_a^{\rm L}$ and $-\hat L_a^{\rm R}$ the su(2) generators acting on ${\cal F}$
and ${\cal F}^*$ respectively. The geometry of the round fuzzy sphere
is then fixed by specifying the Laplacian to be
\eqa
\label{S2Laplacian}
\Delta&=&\hat{\cal L}_a\hat{\cal L}_a\\
&=& {(\hat L_a \hat L_a)}^{\rm L}\otimes \ID+\ID\otimes {(\hat L_a \hat L_a)}^{\rm R}
-2 \hat L_a^{\rm L}\otimes \hat L_a^{\rm R}.
\nonumber
\qea

Let us pause and examine this construction in more detail.
Since, for our purposes, we will not need to diagonalize $\hat L_0$ it is more convenient to work
in a basis where the $SU(2)$ symmetry is manifest. Our preferred basis for ${\cal F}_L$ is the set of vectors
\eq
|\avec\rangle=
|\alpha_1,\cdots,\alpha_L\rangle=\frac{1}{\sqrt{L!}}
a^\dagger_{\alpha_1}\cdots a^\dagger_{\alpha_L}|0\rangle \ ,
\qe
which satisfy the orthogonality relation
\eq
\langle\bvec|\avec\rangle={\cal S}_{{\overline{\bvec}}\avec}={\cal S}^{\bvec}_{\avec}
\qe
or more explicitly
\eq
\langle\beta_1,\cdots,\beta_L|\alpha_1,\cdots,\alpha_L\rangle=
{\cal S}^{\beta_1\dots\beta_L}_{\alpha_1\cdots\alpha_L}=
\frac{1}{L!}\delta^{\beta_1}_{\{\alpha_1}\dots\delta^{\beta_L}_{\alpha_L\}}
\qe
where ${\cal S}^{\beta_1\dots\beta_L}_{\alpha_1\cdots\alpha_L}$ is the projector onto
totally symmetric tensors.

A basis for the finite dimensional algebra is provided by a pairing of vectors
from ${\cal F}_L$ and ${\cal F}_L^*$ of the form $\{|\avec\rangle\langle\bvec|\}$ and we denote
the space spanned by this basis as ${\cal F}\otimes{\cal F}^*$
\eq
{\cal F}\otimes{\cal F}^*=span\{|\avec\rangle
\langle\bvec|\}.
\qe
A general matrix is then given by
\eqa
\label{S2_generalmatrix}
{\bf M}&=&\frac{1}{L!}{M^{\alpha_1\cdots\alpha_L}}_{\beta_1\cdots\beta_L}
 a_{\alpha_1}^\dagger \cdots a_{\alpha_L}^\dagger \ket 0 \bra 0 a^{\beta_1} \cdots a^{\beta_L}
\nonumber\\
&&\\
&=&{M^{\avec}}_{\bvec}|\avec\rangle\langle\bvec| \ .
\nonumber \qea
The Laplacian is represented as in (\ref{S2Laplacian}) where, in terms of raising and lowering operators, we have:
\eq
\label{calLpm}
\hat{\cal L}_{\pm}{\bf M}=[\hat L_{\pm},{\bf M}], \qquad \hat{\cal L}_0{\bf M}=[\hat L_0,{\bf M}].
\qe
In particular the matrix
\eq
{\bf \unit} = |\avec\rangle\langle\avec|=|\alpha_1,
\cdots\alpha_L\rangle\langle\alpha_1,\cdots\alpha_L|
\qe
is easily seen to be in the kernel of $\hat {\cal L}_\pm$ and $\hat {\cal L}_0$ and
represents the unit operator. Taking the trace, we have
\eq
\langle\avec|\avec\rangle:=\langle\alpha_1,\cdots\alpha_L|\alpha_L,\cdots\alpha_1\rangle=L+1
\qe
the dimension of the matrix algebra. It is further useful to introduce the notation
\eq
\label{Fockcontractions}
|\avec_l,\gvec_{L-l}\rangle\langle\gvec_{L-l},\bvec_l|
=|\alpha_1,\cdots,\alpha_l,\gamma_{l+1},\cdots,\gamma_L\rangle
\langle\gamma_{L},\cdots,\gamma_{l+1},\beta_l,\cdots\beta_1|
\qe
for basis elements where the $L-l$ indices $\gvec_{L-l}$ are contracted.

The different eigenspaces of the Laplacian provide the polarization
tensors ${{\bf Y}_{\avec_l}}^{\bvec_l}$ with $l=0\dots L$, where all
remaining contractions between $\avec_l$ and $\bvec_l$ have been
removed.  These are easily constructed by Gram-Schmidt
orthogonalization. The polarization tensors then satisfy
\eq
\label{Polarization-tensor-normalization}
{({{\bf Y}_{\avec_l}}^{\bvec_l})}^\dagger= {{\bf Y}_{\bvec_l}}^{\avec_l} \quad {\rm and }\quad
\frac{Tr}{L+1}({{\bf Y}_{\avec_l}}^{\bvec_l}{{\bf Y}_{\svec_{l'}}}^{\tauvec_{l'}})
=\delta_{ll'}{{\cal P}_{{\avec_l},{\svec_{l'}}}}^{\bvec_{l},\tauvec_{l'}}
\qe
where ${{\cal P}_{{\avec_l},{\svec_{l'}}}}^{\bvec_{l},\tauvec_{l'}}$
is the projector onto symmetric traceless tensors, {\it i.e.}\ it removes all
traces between $\avec_l$ and $\bvec_l$ in (\ref{Fockcontractions}). For example
\eq
{\bf Y}=\unit,\qquad {{{\bf Y}}_{\alpha}}^{\beta}=
\sqrt{\frac{6 L}{L+2}}\left(|\alpha,\gvec_{L-1}\rangle\langle\beta,\gvec_{L-1}|
-\frac{1}{2}{\delta_{\alpha}}^{\beta}\unit\right)
\qe
and the polarization tensor, ${{\bf Y}_{\avec_l}}^{\bvec_l}$, for angular momentum $l$
has free indices $\avec_l$ and $\bvec_l$ with eigenvalue $l(l+1)$ for
the Laplacian (\ref{S2Laplacian}) {\it i.e.}\
\eq
\hat {\cal L}^2 {{\bf Y}_{\avec_l}}^{\bvec_l}= l(l+1){{\bf Y}_{\avec_l}}^{\bvec_l} \ .
\qe
The construction has a very clean group theoretical meaning: The decomposition into
polarization tensors is the decomposition of the tensor product representation
into irreducible representations and is expressed
in Young diagrams as
\eqa
\label{fuzzy_sphere_l_content}
\overbrace{\young(\ \Dots \ )}^L
\otimes{\overbrace{\young(\ \Dots \ )}^L}
=
\ID\oplus\ \young(\ \ ) \oplus
\cdots\oplus \overbrace{\young(\ \Dots \Dots \ )}^{2l}
\oplus\cdots\oplus \overbrace{\young(\ \Dots \Dots \ )}^{2L} \ .
\qea
We see that increasing $L$ by one adds one additional multiplet to
(\ref{fuzzy_sphere_l_content}).
Our complete set of polarization tensors is
\eq
\label{projecttoremovetraces}
{{\bf Y}_{\avec_l}}^{\bvec_l}=\frac{\sqrt{L+1}}{\sqrt{Q(l,L)}}
{{{\cal P}_{\avec_l,\svec_l}}}^{\bvec_l,\tauvec_l}
|\tauvec_l,\gvec_{L-l}\rangle\langle\svec_l,\gvec_{L-l}| \ \ ,
\qe
where ${{{\cal P}_{\avec_l,\svec_l}}}^{\bvec_l,\tauvec_l}$ is the projector
onto the irreducible representation
\eqa \overbrace{\young(\ \Dots \Dots \ )}^{2l}\ \ ,
\qea
{\it i.e.}\ it removes all traces associated with contractions of
the free indices of ${\cal F}_L$ and ${\cal F}_L^*$. The coefficient $Q(l,L)$ in the
normalization arises due to the contracted oscillators. With $l=L$ before the application of
the projector ${{{\cal P}_{\avec_l,\svec_l}}}^{\bvec_l,\tauvec_l}$
there are no contracted oscillators and the normalization obtained from (\ref{Polarization-tensor-normalization}) is $\sqrt{L+1}$ so
$Q(L,L)=1$.

The more general normalization can be obtained by observing that the contraction
over the $L-l$ indices $\gvec_{L-l}$ corresponds to the repeated embedding of a matrix
with angular momentum cutoff $k$ into a matrix with cutoff $k+1$ by adding contracted
oscillators, the operation being repeated from $k=l+1$ up to $k=L$.
Thus for ${\bf M}\in {\rm Mat_{k+1}}$,
\eq
a^\dagger_{\gamma_{k+1}}{\bf M}a^{\gamma_{k+1}}\in {\rm Mat_{k+2}}
\qe
and we see from the above discussion that the polarization tensor content has not changed so
an embedded matrix still has angular momentum only up to $k$;
the top angular momentum $l=k+1$ is naturally absent.
Now if we rewrite (\ref{S2Laplacian}) we obtain the relations:
\eqa
\label{updown_mapping_for_s2}
a^\dagger_{\alpha}a^\beta {\bf M}a^\dagger_{\beta}a^{\alpha}
&=&\left(\hat N(\hat N+1)-\hat {\cal L}^2\right){\bf M} \quad {\rm and}
\nonumber\\
&&\\
a^{\alpha}a^\dagger_\beta {\bf M}a^{\beta}a^\dagger_{\alpha}
&=&\left((\hat N+1)(\hat N+2)-\hat {\cal L}^2\right){\bf M} \ ,
\nonumber
\qea
where we used $[\hat N,{\bf M}]=0$. In the first expression in (\ref{updown_mapping_for_s2})
the matrix ${\bf M}$ is reduced before being re-embedded thus projecting out the top multiplet
with $l=k$ from the spectrum of the Laplacian $\hat {\cal L}^2$, while in the second
the matrix ${\bf M}$ is embedded before being reduced.
From these and the fact that only one multiplet is added by increasing the cutoff by one
we can deduce that the eigenvalues of $\hat {\cal L}^2$ are
$l(l+1)$. Also the factor $Q(l,L)$ in the normalization of polarization tensors is
obtained from (\ref{Polarization-tensor-normalization}) and given by
\eq
Q(l,L)=\frac{{(l!)}^2}{(2l+1)!}\frac{(L-l)!(L+l+1)!}{{(L!)}^2} \ ,
\qe
with $Q(L,L)=1$.

It is now easy to relate this formulation to one in terms of functions.
A given matrix ${\bf M}\in {\rm Mat_{L+1}}$ can be expanded in polarization tensors,
\eq
\label{expansion_of_square_matrix}
{\bf M}=\sum_{l=0}^{L}{M^{\avec_l}}_{\bvec_l}{{\bf Y}_{\avec_l}}^{\bvec_l} \ .
\qe
Define the {\it symmetric symbol density} as the matrix ${\bf \rho}_L({\bar z}, z)$
\eq
\label{symmetric_map_to_functions}
{\bf \rho}_L({\bar z},z)=\sum_{l=0}^{L}{Y_{\bvec_l}}^{\avec_l}({\bar z}, z){{\bf Y}_{\avec_l}}^{\bvec_l}
\qe
where
\eq
\label{Ylms}
{Y_{\bvec_l}}^{\avec_l}({\bar z}, z)=\frac{\sqrt{(2l+1)!}}{l!}
{{{\cal P}_{\bvec_l,\svec_l}}}^{\avec_l,\tauvec_l}
{\bar z}_{\tau_1}\dots {\bar z}_{\tau_l}
{z}^{\sigma_1}\dots {z}^{\sigma_l}
\qe
are the ordinary spherical harmonics in a spinorial basis normalized such that
\eq
\label{Spherical-harmonics-normalization}
\frac{1}{{\rm Vol}(S^2)}\int_{S^2}\omega\ (
{{Y}_{\bvec_l}}^{\avec_l}{{ Y}_{\tauvec_{l'}}}^{\svec_{l'}})
=\delta_{ll'}{{\cal P}_{\bvec_{l},\tauvec_{l'}}}^{{\avec_l},{\svec_{l'}}} \ ,
\qe
with $\omega$ the volume form and ${\bar z}_{\alpha}z^{\alpha}=1$.

Then the trace
\eq
\label{maptofunctionss2}
M({\bar z}, z)=\frac{Tr}{L+1}\left({\bf \rho}_L({\bar z}, z){\bf M}\right)
\qe
gives a function on $S^2$ whose expansion in spherical harmonics has the same coefficients
as the coefficients of the matrix in terms of the polarization tensors.
We can approximate a function $f({\bar z},z) \in C^{\infty}(S^2)$ by the matrix
\eq
\label{map_to_matrices}
{\bf M}_f=\frac{1}{{\rm Vol}(S^2)}\int_{S^2}\omega {\bf \rho}_L f \ ,
\qe
which is a matrix whose coefficients in an expansion in polarization tensors
coincides with $f$ up to angular momentum $L$ and all higher coefficients
are projected to zero. If one substitutes the function $M({\bar z},z)$ obtained
from ${\bf M}$ in (\ref{maptofunctionss2}) into (\ref{map_to_matrices}) one recovers
the matrix ${\bf M}$.

An equally good map to functions is provided by\footnote{The conventions here are set by
noting that for $L=1$ we have
$D^{\frac{1}{2}}_{\alpha,\frac{1}{2}}(z,{\bar z})=z^{\alpha}$
and $|z,1\rangle={z}^{\alpha}|\alpha\rangle={z}^{\alpha}a^\dagger_{\alpha}|0\rangle
={({\bar z}_{\alpha}a^\alpha)}^\dagger|0\rangle$,
where $D^{j}_{m_1,m_2}(g)$ are the
Wigner D-matrices for $SU(2)$, and on the entire Fock space, ${\cal F}$,
coherent states are defined as eigenvectors of the annihilation operator. }
the diagonal coherent state map \cite{Perelomov:1986tf}
\eq
\label{coherent_State_map}
M_L({\bar z},z)=\langle z,L|{\bf M}|L,z\rangle \ ,
\qe
where we have taken the trace of
\eqa
\label{coherent_state_map_to_functions}
|z,L\rangle\langle L,z|
&:=&\frac{1}{L!}{(z^{\alpha}a^\dagger_{\alpha})}^L|0\rangle\langle0|{({\bar z}_{\beta}a^{\beta})}^L \\
&=&\sum_{l=0}^{L} \frac{T_L^{1/2}(l)}{L+1}{Y_{\bvec_l}}^{\avec_l}({\bar z}, z)
{{\bf Y}_{\avec_l}}^{\bvec_l}
\ ,
\nonumber\qea
which effects the simple replacement $a^\alpha\rightarrow z^\alpha$, $a^\dagger_{\alpha}\rightarrow {\bar z}_\alpha$ in (\ref{S2_generalmatrix}) and the removal of $1/L!$.

The principal difference in the two maps to functions (\ref{symmetric_map_to_functions}) and
(\ref{coherent_state_map_to_functions})
is the presence in the latter of the coefficients
\eq
\label{T_L_S2}
T_L(l)=\frac{L!(L+1)!}{(L-l)!(L+l+1)!}={\left(1-\frac{l(l+1)}{L(L+1)}\right)}^{-1}T_{L-1}(l)\ ,
\qe
which alter the coefficients in the expansion of a function from those in the
expansion of the corresponding matrix.
Note that for $L\rightarrow\infty$ with $l$ fixed $T_L(l)\rightarrow 1$ so the
distortion disappears as $L\rightarrow\infty$.  The image functions $M_L$ and $M$
are related by
\eq
\label{density_mapping_operator}
M_L({\bar z}, z)={\cal T}_L^{1/2}({\cal L}^2) M({\bar z},{ z})\ ,
\qe
where the rotationally invariant operator ${\cal T}_L({\cal L}^2)$ has eigenvalues
$T_L(l)$ \cite{Berezin:1974du,Berezin:1975:Izv,Dolan:2001gn}.
The principal advantage of the coherent state map is the simplicity of the
associated *-product \cite{Balachandran:2001dd,Kurkcuoglu:2006iw}.
In contrast that induced by (\ref{symmetric_map_to_functions}) is more complicated
but its leading term in a large $L$ expansion is the Poisson bracket, {\it i.e.}\ the
symmetric part of the *-product vanishes in the leading term.

There is an alternative set of operators
${\tilde{\cal L}}_{\pm}$ to those in (\ref{calLpm}) obtained
by interchanging both the roles of left and right and $+$ and $-$ to obtain
new operators:
\eq
\tilde{\cal L}_{\pm}{\bf M}=-[\hat L_{\mp},{\bf M}], \qquad \tilde{\cal L}_0{\bf M}=-[\hat L_0,{\bf M}] \ .
\qe
These generators can be induced naturally by observing that the set of states is unchanged
by transforming to the oscillators
\eq
\label{A_alpha_bar_for_s2}
a_{\alpha}:=a^\beta\epsilon_{\beta\alpha}\equiv a^{\overline{\alpha}} \ ,
\qe
{\it i.e.}\ substituting $a^2\rightarrow -a_1$ and $a^1\rightarrow a_2$.
The set $a^\dagger_{\overline{\alpha}}$ generate precisely the same matrix algebra
and with this substitution, we have
\eqa
{\tilde L}_{+}={(a_1)}^\dag a_2=-a_2^\dagger a^1=-\hat L_{-} ,
\qquad {\tilde L}_{-}={(a_2)}^\dag a_1=-a_1^\dagger a^2=-\hat L_{+}\\
{\tilde L}_{0}=\frac{1}{2}({(a_1)}^\dag a_1-{(a_2)}^\dagger a_2)
=\frac{1}{2}(a_2^\dagger a^2-a_1^\dagger a^1)
=-\hat L_0
\qea
and $\tilde {\cal L}^2=\hat {\cal L}^2$ so the resulting Laplacian is however unchanged.
This reflects the fact that complex conjugating a
representation gives a unitarily equivalent one for $su(2)$.

There is yet a further realization of the Laplacian. This was first given in
\cite{Grosse:1995jt} in terms of the
operators:\footnote{One can multiply both $\hat K_{+}$ and
$\hat K_{-}$ by opposite phases to get equivalent operators.}
\eq
\begin{array}{lcllll}
\label{Ks_as_module_maps}
\hat K_{+}:=&{(a^\dagger_{\alpha})}^{\rm L} {({(a^\dagger)}^{\alpha})}^{\rm R}&:& {\cal F}_{L}\otimes {\cal F}_{L}^*
&\longmapsto& {\cal F}_{L+1}\otimes {\cal F}_{L-1}^*\\
\hat K_{-}:=&{(a^\alpha)}^{\rm L} {(a_\alpha)}^{\rm R}&:&{\cal F}_{L}\otimes {\cal F}_{L}^*
&\longmapsto& {\cal F}_{L-1}\otimes {\cal F}_{L+1}^*\\
\hat K_{0}:=&\frac{1}{2}(\hat N^{\rm L}-\hat N^{\rm R})&:&{\cal F}_{L}\otimes {\cal F}_{L}^*
&\longmapsto& 0
\end{array}
\qe
where the $\rm L$ and $\rm R$ superscripts indicate that the operators act on the
left or right, {\it i.e.}\ on ${\cal F}_L$ or ${\cal F}_L^*$, respectively.

Note: These operators do not require that the left and right truncated Fock spaces have
the same dimension, more generally, they act on bimodules of the form
${\cal F}_{n_{\rm L}}\otimes {\cal F}_{n_{\rm R}}^*$ which represent non-square matrices which are
left modules for the algebra ${\rm Mat_{n_{\rm L}+1}}$ and right modules for ${\rm Mat_{n_{\rm R}+1}}$.
For our purposes, it is convenient to denote such a generic element of a left module of ${\rm Mat_{L+1}}$
by ${\bf M}_q$, where ${\bf M}_q\in {\cal F}_L\otimes {\cal F}_{L-q}^*$
and we have
\eqa
\label{su2KsDefined}
\hat K_{+}{\bf M}_q&=&\epsilon^{\alpha\beta}a^\dagger_\beta{\bf M}_q a^\dagger_\alpha \ ,
\nonumber\\
\qquad \hat K_{-}{\bf M}_q&=&\epsilon_{\alpha\beta}a^{\beta}{\bf M}_q a^{\alpha} \ ,
\\
\hat K_0{\bf M}_q&=&\frac{1}{2}[\hat N,{\bf M}_q]=\frac{q}{2}{\bf M}_q.
\nonumber\qea
We see that $\hat K_{+}$ and $\hat K_{-}$ change the module structure
as in (\ref{Ks_as_module_maps})
while $\hat K_0$ measures the non-squareness of a given bimodule
and in particular, the matrix algebra can be identified with the kernel of $\hat K_0$
({\it i.e.}\ with $q=0$, ${\bf M}\in {\rm Mat_{L+1}}$ is in the kernel of $\hat K_0$).
Repeated applications of $\hat K_\pm$ map us further along the sequence of modules.

Although  $\hat K_+$ and $\hat K_-$ take us out of the algebra of square matrices
the products $\hat K_+\hat K_-$ and $\hat K_-\hat K_+$ do not.  In general, one
obtains
\eqa
\hat K_+\hat K_-{\bf M}_q&=&\hat N{\bf M}_q(\hat N+1)-
a^\dagger_\alpha a^{\beta}{\bf M}_q a^\dagger_{\beta}a^{\alpha}
\\
\hat K_-\hat K_+{\bf M}_q&=&(\hat N+1){\bf M}_q\hat N-
a^\dagger_\alpha a^{\beta}{\bf M}_q a^\dagger_{\beta}a^{\alpha}
\nonumber
\qea
and the operators $\hat K_{\pm}$ and $\hat K_0$ are easily seen to
satisfy the $su(2)$ commutation relations.
Similar manipulations for the Laplacian (\ref{S2Laplacian})
acting on ${\bf M}_q$ yield
\eq
\label{S2_universal_Laplacian_for_line_bundles}
\hat{\cal L}^2{\bf M}_q =(L-\frac{q}{2})(L-\frac{q}{2}+1){\bf M}_q
-a^\dagger_{\alpha}a^{\beta}{\bf M}_q a^\dagger_{\beta}a^{\alpha}.
\qe
Furthermore, we have
\eqa
\label{Laplacian_in_terms_of_Ks}
\hat K^2{\bf M}_q&=&\frac 1 2 \bigl( \hat K_-\hat K_+ + \hat K_+\hat K_- +2\hat K_0^2\bigr){\bf M}_q\\
\nonumber
&=&(L-\frac{q}{2})(L-\frac{q}{2}+1) {\bf M}_q
-a^\dagger_\alpha a^{\beta}{\bf M}_q a^\dagger_{\beta}a^{\alpha}.
\qea
Hence the potential Laplacians $\hat {\cal L}^2$, $\tilde{\cal L}^2$ and $\hat K^2$
are all equal and we are left with a unique option
for the round Laplacian $\Delta=\hat {\cal L}^2=\hat K^2$ on $S^2_F$ given by
(\ref{S2_universal_Laplacian_for_line_bundles}).

As pointed out in \cite{Grosse:1995jt} these latter non-square matrices capture topologically
nontrivial field configurations on the fuzzy sphere and can be taken to be the noncommutative
versions of holomorphic line bundles, with the eigenvalue of $\hat K_0$ given by $q/2$ and $q$
counting the winding number, so that $q> 0$ describe $\overline {\cal O}(q)$ bundles and $q<0$
describe ${\cal O}(-q)$ bundles. The Laplacian for these line bundles is naturally given by
(\ref{Laplacian_in_terms_of_Ks}) while that based on the construction given in
\cite{Baez:1998he,Balachandran:1999hx,Grosse:2001qt,CarowWatamura:2004ct} is more cumbersome.

The generalization of (\ref{expansion_of_square_matrix})
to non-square matrices\footnote{Polarization tensors for non-square matrices in the standard basis were
constructed by Pre\v{s}najder in \cite{Presnajder:1999ky} and are readily
related to those we present. }
${\bf M}_q$ is given by
\eq
\label{D_polarization_expansion}
{\bf M}_q=\sum_{l=0}^{L-q}{M^{\avec_{l+q}}}_{\bvec_{l}}{{\bf D}_{\avec_{l+q}}}^{\bvec_{l}}
\qe
and ${{\bf D}_{\avec_{l+q}}}^{\bvec_{l}}$ are the polarization tensors of $spin-(l+\frac{q}{2})$
which are obtained by projection onto the relevant representation in the decomposition of
${\cal F}_L\otimes{\cal F}_{n_{\rm R}}^*$ where $n_{\rm R}=L-q$; {\it i.e.}\
\eq
\label{D_polarization_tensors}
{{\bf D}_{\avec_{l+q}}}^{\bvec_{l}}=\frac{\sqrt{n_{\rm R}+1}}{\sqrt{Q(l,q,L)}}
{{{\cal P}_{\avec_{l+q},\svec_{l}}}}^{\bvec_{l},\tauvec_{l+q}}
|\tauvec_{l+q},\gvec_{n_{\rm R}-l}\rangle\langle\svec_{l},\gvec_{n_{\rm R}-l}|
\qe
as in (\ref{projecttoremovetraces}) above and normalized such that
\eq
\frac{1}{n_{\rm R}+1}Tr\left({({{\bf D}_{\avec_{l+q}}}^{\bvec_{l}})}^\dagger
{{\bf D}_{\svec_{l'+q'}}}^{\tauvec_{l'}}\right)
 =\delta_{ll'}\delta_{qq'}{{{\cal P}_{\bvec_{l},\svec_{l+q}}}}^{\avec_{l+q},\tauvec_{l}}
\qe
\noindent
and
\eq
Q(l,q,L)=\frac{l!(l+q)!}{(2l+1+q)!}\frac{(L-q-l)!(L+l+1)!}{(L-q)!L!}.
\qe
The Laplacian $\Delta$ is diagonal on the polarization tensors and we
have
\eq
\Delta {{\bf D}_{\avec_{l+q}}}^{\bvec_{l}}
=(l+\frac{q}{2})(l+\frac{q}{2}+1){{\bf D}_{\avec_{l+q}}}^{\bvec_{l}}
\ , \quad {\rm with }\quad l=0,\dots L-q \ .
\qe

The {\it symbol density} analogous to (\ref{symmetric_map_to_functions}) which now
provides a map to equivariant sections of line bundles over $S^2$ is
\eqa
\label{non_square_matrix_symmetric_map_to_functions}
{\bf \rho}_{L,q}({\bar z},{ z})&=&
\sum_{l=0}^{L-q}{D_{\bvec_{l+q}}}^{\avec_{l}}({\bar z},z)
{({{\bf D}_{\bvec_{l+q}}}^{\avec_{l}})}^\dagger\nonumber\\
&=&\sum_{l=0}^{L-q}{D_{\bvec_{l+q}}}^{\avec_{l}}({\bar z},z){{{\bf D}_{\avec_{l}}}^{\bvec_{l+q}}}\ ,
\qea
where
\eq
\label{Wigner-D-spin-basis}
{D_{\avec_{l+q}}}^{\bvec_{l}}({\bar z},{ z})=
\sqrt{\frac{(2l+1+q)!}{l!(l+q)!}}
{{{\cal P}_{\avec_{l+q},\svec_{l}}}}^{\bvec_{l},\tauvec_{l+q}}\bar z_{\tau_1}\dots\bar z_{\tau_{l+q}}z^{\sigma_1}\dots z^{\sigma_l}
\equiv {\overline D}^{j}_{m,\frac{q}{2}}(z,{\bar z})
\qe
with $j=l+\frac{q}{2}$ and $D^{j}_{m,s}(z,{\bar z})$ are the Wigner D-matrices.

The relevant coherent state map for matrices ${\bf M}_q$ is provided by
\eqa
\label{m_q_coherentstatemap}
|z,n_{\rm R}\rangle\langle n_{\rm L},z|&:=&\frac{1}{\sqrt{n_{\rm R}!n_{\rm L}!}}
{(z^{\alpha}a^\dagger_{\alpha})}^{n_{\rm R}}|0\rangle
\langle0|{(\bar z_{\alpha}a^{\alpha})}^{n_{\rm L}}
\nonumber\\
&=&\sum_{l=0}^{n_{\rm R}} \frac{T^{1/2}_{L}(l,q)}{n_{\rm R}+1}
{D_{\bvec_{l+q}}}^{\avec_{l}}({\bar z},{ z}){({{\bf D}_{\bvec_{l+q}}}^{\avec_{l}})}^\dagger
\qea
where $n_{\rm L}=L$ and $n_{\rm R}=L-q$ and
\eq
T_{L}(l,q)
=\frac{L!(L-q+1)!}{(L-q-l)!(L+l+1)!} \ .
\qe

If we use the diagonal coherent state map (\ref{m_q_coherentstatemap})
it is easy to establish the correspondence
\eqa
\label{induced_identification_cs}
{(\frac{1}{\sqrt{\hat N}}a^\dagger_{\alpha})}^{\rm L}\longmapsto {\bar z}_{\alpha}\qquad
{(a^{\alpha}{\sqrt{\hat N}})}^{\rm L}\longmapsto \dd{}{\bar z}_\alpha
\nonumber\\
\\
{(a^{\alpha}\frac{1}{\sqrt{\hat N}})}^{\rm R}\longmapsto {z}^{\alpha}\qquad
{({\sqrt{\hat N}}a^\dagger_{\alpha})}^{\rm R}\longmapsto \dd{}{z}{}_\alpha \ .
\nonumber
\qea
\noindent
The presence of $\sqrt{\hat N}$ in these expressions takes care of the normalization
of the states.

Now that we have mapped matrices to functions and  non-square matrices
to equivariant sections of monopole bundles we are in a position
to map the various differential operators and the Laplacian to their commutative analogues.
However, for our later discussion it is useful to pause and make a small digression back
to the commutative setting and review the relevant operators there.

Using the identification induced by (\ref{induced_identification_cs}) and the operators
identified in appendix {\bf \ref{commutative_s2}} we have
\eq
\hat L^{\rm L}_+={(a_1^\dagger a^2)}^{\rm L}\longmapsto  {\bar z}_1\frac{\partial}{{\partial {\bar z}}_2}
=- \bar L_-\ ,\quad
\hat L^{\rm R}_+={(a_1^\dagger a^2)}^{\rm R}\longmapsto {z}^2\frac{\partial}{\partial { z}^1}=-  L_+\ ,
\qe
\eq
\hat L^{\rm L}_-={(a_2^\dagger a^1)}^{\rm L}\longmapsto {\bar z}_2\frac{\partial}{{\partial \bar z}_1}
=- \bar L_+ \ ,\quad
\hat L^{\rm R}_-={(a_2^\dagger a^1)}^{\rm R}\longmapsto {z}^1\frac{\partial}{\partial { z}^2}=-L_-\ ,
\qe
\eq
\hat L^{\rm L}_{0}=\frac{1}{2}\left({(a_1^\dagger a^1)}^{\rm L}-{(a_2^\dagger a^2)}^{\rm L}\right)
\longmapsto
\frac{1}{2}\left( \bar z_1\frac{\partial}{\partial {\bar z}_1}
-{ \bar z}_2\frac{\partial}{\partial \bar z_2}\right)=- \bar L_0\ ,
\qe
\eq
\hat L^{\rm R}_{0}=\frac{1}{2}\left({(a_1^\dagger a^1)}^{\rm R}-{(a_2^\dagger a^2)}^{\rm R}\right)
\longmapsto
\frac{1}{2}\left({z}^1\frac{\partial}{\partial { z}^1}
-{ z}^2\frac{\partial}{\partial z^2}\right)=- L_0 \ .
\qe
so the operators $\hat {\cal L}_a$ map to the right invariant
vector fields ${\cal L}_a$ as:
\eqa
\nonumber\\
\hat{\cal L}_{+}\longmapsto {\cal L}_{+}&=&  L_+ - \bar L_- \ ,
\nonumber\\
\hat{\cal L}_{-}\longmapsto {\cal L}_{-}&=&  L_- - \bar L_+ \ , \\
\hat{\cal L}_{0}\longmapsto {\cal L}_{0}
&=& L_0 - \bar L_0\  .
\nonumber\qea

Similarly using the coherent state map (\ref{m_q_coherentstatemap}) and equation (\ref{rightStwo}) we find
\eqa
\label{KtoR_via_coherent_state_map}
\hat K_{+}&\longmapsto& -\sqrt{\frac{L+1}{L-q}} {\cal R}_- \ ,\\
\hat K_{-}&\longmapsto &-\sqrt{\frac{L-q+1}{L}}{\cal R}_+ \ ,\\
\hat K_{0}&\longmapsto &\quad -{\cal R}_0 \ ,
\qea
where the normalizations on the right do not affect the $su(2)$ commutation relations.
These factors are harmless and could be removed by replacing
\eq
\hat K_+\longmapsto \frac{1}{\sqrt{\hat N^{\rm L}}}\hat K_{+}{\sqrt{\hat N^{\rm R}}}\quad {\rm and}\quad
\hat K_-\longmapsto \frac{1}{\sqrt{\hat N^{\rm R}}}\hat K_{-}{\sqrt{\hat N^{\rm L}}} \ .
\qe
and $\hat K_0$ is unaffected.
Finally with the above, we have of course $\Delta\longmapsto -\nabla^2$.

We could alternatively have built our map to functions
using $a^\dagger_{\overline{\alpha}}$ {\it i.e.}\ via
\eq
\ket{\bar z,n_{\rm R}}\bra{z,n_{\rm L}}=
\frac{1}{\sqrt{n_{\rm R}!n_{\rm L}!}}(\bar z_\alpha a^\dagger_{\overline{\alpha}})^{n_{\rm R}}
\ket{0}\bra{0}( z^\alpha a_\alpha)^{n_{\rm L}}\ .
\qe
This would end up in replacing $z$ by ${\bar z}$, or equivalently
replacing ${\bf M}$ with ${\bf M}^\dagger$
in the map between matrices and functions.

A principal result of this paper is the generalization of the
operators $\hat K_{\pm}$ to the case of $\CP^N$ and we will see that it
will give us access to projective modules that
correspond to the algebra of functions tensored with holomorphic line bundles
and complex vector bundles over $\CP^N$.

\section{Fock space construction of $\CP^N_F$}
\label{CPN-Fuzzy}

The above discussion is very easily adapted to $\CP^N$ by considering $a^{\alpha}$,
a multiplet of
$N+1$ oscillators. In this case the construction analogous
to (\ref{u2SchwingerJordanops}) and (\ref{u2SchwingerJordanalg})
gives the $u(N+1)$ algebra based on the operators
\eq
\hat N=a^\dagger a, \qquad {\rm and }\qquad \hat L_a=a^\dagger \frac{\lambda_a}{2}a
\qe
where $\lambda_a$ are the Gell-Mann matrices\footnote{In our conventions the Gell-Mann
matrices in the anti-fundamental are denoted $\overline{\lambda}_a$ and given by
\eq
{({\overline\lambda}_a)}_{\alpha {\overline\beta}}
=-{({\lambda_a})}_{{\overline{\beta}\alpha}}.
\qe
}
of $SU(N+1)$ with $a$ running from $1$ to $(N+1)^2-1$.
We can again decompose the Fock space, ${\cal F}$, generated freely by $a^\dagger_{\alpha}$,
into a direct sum of finite dimensional spaces
${\cal F}_L$ carrying the  irreducible representation corresponding to
$L$-fold symmetric tensor product of the fundamental of $su(N+1)$, which
we labeled by the eigenvalue of $\hat N$.

We similarly define states spanning ${\cal F}_L$ to be
\eq
|\alpha_1,\cdots\alpha_L\rangle=\frac{1}{\sqrt{L!}}a^\dagger_{\alpha_1}\cdots a^\dagger_{\alpha_L}|0\rangle.
\qe
The unit matrix is represented by
\eq
{\bf \unit} = |\alpha_1,\cdots\alpha_L\rangle\langle\alpha_1,\cdots\alpha_L|
\qe
and has trace
\eq
\langle\alpha_1,\cdots\alpha_L|\alpha_1,\cdots\alpha_L\rangle=d_N(L)=\frac{(N+L)!}{N!L!}\ ,
\qe
where $d_N(L)$ is the dimension of the $L$-fold symmetric tensor product representation of
$SU(N+1)$. A generic matrix ${\bf M}$ is as in the case of $S_F^2$ given by
\eq
\label{general_matrix_cpn}
{\bf M}=\frac{1}{L!}{M^{\alpha_1\cdots\alpha_L}}_{\beta_1\cdots\beta_L}
 a_{\alpha_1}^\dagger \cdots a_{\alpha_L}^\dagger \ket 0 \bra 0 a^{\beta_1} \cdots a^{\beta_L}.
\qe

The geometry of $\CP^N_F$ can then be specified by building the derivatives
$\hat{\cal L}_a$, as we did for $S^2_F$, from the commutator action
\eq
\hat {\cal L}_a{\bf M}=[\hat L_a,{\bf M}], \qquad
{\rm where}\qquad \hat L_a=a^\dagger \frac{\lambda_a}{2}a
\qe
with $a$ running from $1$ to $(N+1)^2-1$. The Laplacian given by the quadratic Casimir
$\Delta=\hat {\cal L}_a \hat {\cal L}_a$, can be expressed
in the form (\ref{S2_universal_Laplacian_for_line_bundles}) as
\eq
{\hat{\cal L}}^2{\bf M}
=\left(L(L+N){\bf M}-a^\dagger_{\alpha}a^\beta {\bf M}a^\dagger_{\beta}a^{\alpha}\right) \ .
\qe
This is easily obtained using
\eq
\label{gell-mann_completeness}
{(\lambda_{a})}_{\overline{\alpha}{\beta}} {(\lambda_{a})}_{\overline{\mu}{\nu}} =
2 \delta_{\overline{\alpha}\nu}\delta_{\overline{\mu}\beta}
-\frac{2}{N+1}\delta_{\overline{\alpha}\beta}\delta_{\overline{\mu}\nu} \ ,
\qe
yielding
\eq
2\hat L_a^{\rm L}\otimes \hat L_a^{\rm R}={(a_{\alpha}^\dagger a^{\beta})}^{\rm L}\otimes {(a^\dagger_{\beta} a^{\alpha})}^{\rm R}
-\frac{\hat N^{\rm L}\otimes\hat N^{\rm R}}{N+1}
\qe
and
\eq
\hat C_2=\hat L_a^{\rm L} \hat L_a^{\rm L}= \frac{N}{2(N+1)}\hat N(\hat N+N+1).
\qe
More generally, we can write
\eq
\label{cpn_universal_Laplacian_for_line_bundles}
\hat {\cal L}^2=
\hat C_2\otimes \ID+\ID\otimes \hat C_2+\frac{\hat N\otimes\hat N}{N+1}
-{(a_{\alpha}^\dagger a^{\beta})}\otimes {(a^\dagger_{\beta} a^{\alpha})} \ .
\qe
From this we see the analogues of (\ref{updown_mapping_for_s2})
become
\eqa
\label{updown_mapping_for_cpn}
a^\dagger_{\alpha}a^\beta {\bf M}a^\dagger_{\beta}a^{\alpha}
&=&\left(L(L+N)-\hat{\cal L}^2\right){\bf M}
\nonumber\\
&&\\
a^{\alpha}a^\dagger_\beta {\bf M}a^{\beta}a^\dagger_{\alpha}
&=&\left((L+1)(L+1+N)-\hat{\cal L}^2\right){\bf M} \ .
\nonumber
\qea

As in the case of $S^2_F$ the polarization tensors are
given by the decomposition of the tensor product ${\cal F}_L\otimes {\cal F}_L^*$, but now
into irreducible representations of $su(N+1)$. Since ${\cal F}_L$ and ${\cal F}_L^*$ carry
the $L$-fold symmetric tensor product representations of the fundamental and
anti-fundamental of $su(N+1)$, respectively, the relevant group theory decomposition is
\eqa
\overbrace{\young(\ \Dots \ )}^L \,\otimes\,\,
{\overbrace{\young(\ \Dots\ ,\Dots \Dots\Dots ,\ \Dots\ )}^L }
=
\ID\quad\oplus\quad\young(\ \ ,\Dots,\ )
\quad\cdots
\quad\oplus\quad \overbrace{\young(\ \Dots \ \ \Dots \ ,\Dots \Dots \Dots ,\ \Dots \ )}^{2L}
\label{cpn_FFdual_decomposition}
\qea
and the decomposition into polarization tensors is a realization of this decomposition
where the polarization tensors with $2l$-free indices denoted ${{\bf Y}_{\avec_l}}^{\bvec_l}$.
In analogy with (\ref{projecttoremovetraces}), the ${{\bf Y}_{\avec_l}}^{\bvec_l}$ for fuzzy $\CP^N$
are given by
\eq
\label{projecttoremovetraces_CPN}
{{\bf Y}_{\avec_l}}^{\bvec_l}
=\frac{\sqrt{d_N(L)}}{\sqrt{Q_N(l,L)}} {{{\cal P}_{\avec_l,\svec_l}}}^{\bvec_l,\tauvec_l}|\tauvec_l,\gvec_{L-l}\rangle\langle\svec_l,\gvec_{L-l}|
\qe
where
\eq
Q_N(l,L)={\left(\frac{l!}{L!}\right)}^2\frac{(L-l)!(L+l+N)!}{(2l+N)!}
\qe
and ${{{\cal P}_{\avec_l,\svec_l}}}^{\bvec_l,\tauvec_l}$ is the projector onto
the representation
\eq
\overbrace{\young(\ \Dots \ \ \Dots \ ,\Dots \Dots \Dots ,\ \Dots \ )}^{2l}\quad .
\qe
The projector removes all traces corresponding to the lower dimensional representations
in (\ref{cpn_FFdual_decomposition}).

Explicitly, ${\bf Y}=\unit$ for $l=0$  and for $l=1$ we have:
\eq
{{\bf Y}_{\alpha}}^{\beta}=\sqrt{\frac{(N+1)(N+2)L}{L+N+1}}
\left(|\avec,\gvec_{L-1}\rangle\langle\bvec,\gvec_{L-1}|-\frac{1}{N+1}{\delta_{\alpha}}^{\beta}\unit
\right)\ .\qe

The decomposition of a matrix
${\bf M}\in {\rm Mat_{d_N(L)}}$ in terms of polarization tensors is as before
\eq
\label{image_of_M}
{\bf M}=\sum_{l=0}^{L}{M^{\avec_l}}_{\bvec_l}{{\bf Y}_{\avec_l}}^{\bvec_l}
\qe
and the map to functions given by
\eq
M({\bar z}, z)=\frac{Tr}{d_N(L)}({\bf \rho}_L({\bar z},z){\bf M})
\qe
with the symmetric symbol density, ${\bf \rho}_L({\bar z},z)$, given by
\eq
{\bf \rho}_L({\bar z},z)=\sum_{l=0}^{L}{Y_{\bvec_l}}^{\avec_l}({\bar z},z){{\bf Y}_{\avec_l}}^{\bvec_l}
\qe
and the normalizations are the generalization to $\CP^N$ of
(\ref{Polarization-tensor-normalization}) and (\ref{Spherical-harmonics-normalization}),
{\it i.e.}\
\eq
\label{CPN-Polarization-tensor-normalization}
\frac{Tr}{d_N(L)}({{\bf Y}_{\bvec_l}}^{\avec_l}{{\bf Y}_{\tauvec_{l'}}}^{\svec_{l'}})
=\delta_{ll'}{{\cal P}_{{\bvec_l},{\tauvec_{l'}}}}^{\avec_{l},\svec_{l'}}
\qe
and
\eq
\label{CPN-Spherical-harmonics-normalization}
\frac{1}{{\rm Vol}(\CP^N))}\int_{\CP^N}\omega^N (
{{Y}_{\bvec_l}}^{\avec_l}{{ Y}_{\tauvec_{l'}}}^{\svec_{l'}})
=\delta_{ll'}{{\cal P}_{{\bvec_l},{\tauvec_{l'}}}}^{\avec_{l},\svec_{l'}} \ ,
\qe
where $\omega^N$ is the volume form on $\CP^N$ and
\eq
\label{CPN-Ylms}
{Y_{\bvec_l}}^{\avec_l}({\bar z}, z)=\frac{\sqrt{(2l+N)!}}{l!\sqrt{N!}}
{{{\cal P}_{\bvec_l,\svec_l}}}^{\avec_l,\tauvec_l}
{\bar z}_{\tau_1}\dots {\bar z}_{\tau_l}
{z}^{\sigma_1}\dots {z}^{\sigma_l} \ .
\qe
Conversely, the function $f({\bar z},z) \in C^{\infty}(\CP^N)$ is approximated by the matrix
\eq
\label{M_f_approx}
{\bf M}_f=\frac{1}{{\rm Vol}(\CP^N)}\int_{\CP^N}\omega^N {\bf \rho}_L f \ .
\qe
Mapping ${\bf M}_f$ in (\ref{M_f_approx}) back to functions using (\ref{image_of_M})
will result in an approximation to the function $f$ where the coefficients
of all representations of $SU(N+1)$, in $f$,  that lie
above the cutoff representation of dimension $d_N(L)$ are projected to zero.

Again the corresponding coherent state is provided by
\eqa
\label{CPNcoherentstatemap}
|z,L\rangle\langle L,z|&=&\sum_{l=0}^{L} \frac{T_L^{1/2}(l,N)}{d_N(L)}{Y_{\bvec_l}}^{\avec_l}({\bar z},z)
{{\bf Y}_{\avec_l}}^{\bvec_l}
\\
&=&\frac{1}{L!}{({z}^{\alpha}a^\dagger_{\alpha})}^L|0\rangle\langle0|{({\bar z}_{\alpha}a^{\alpha})}^L \ .
\nonumber
\qea
With (\ref{updown_mapping_for_cpn}), the eigenvalues of the $SU(N+1)$ invariant operator
${\cal T}_L({\cal L}^2,N)$ generalizing (\ref{density_mapping_operator})
to the case of $\CP^N$ are found to be
\eq
T_L(l,N)=\frac{L!(L+N)!}{(L-l)!(L+l+N)!} \ .
\qe

More generally, for non-square matrices ${\cal F}_L\otimes {\cal F}_{L-q}^*$ the relevant
group theory decomposition is given by
\eq
\label{cpn_non_square_matrix_decomposition}
\overbrace{\young(\ \Dots \Dots \ )}^L\ \otimes\
{\overbrace{\young(\ \Dots\ ,\Dots \Dots\Dots ,\ \Dots\ )}^{L-q} }
=
\overbrace{\young(\ \Dots \ )}^q\ \oplus\ \overbrace{\young(\ \ \Dots \  ,\Dots,\ )}^{q+2}
\ \cdots
\ \oplus\ \overbrace{\young(\ \Dots \ \ \Dots \ \Dots \ ,\Dots \Dots \Dots ,\ \Dots \ )}^{2L-q}
\qe
which in terms of Dynkin indices reads
\eq
(L,\dots,0,0)\otimes(0,0,\dots,L-q)=
{\mbox{\scriptsize$\displaystyle{\boplus^{L-q}_{l=0}}$}}\ (l+q,0,\dots,0,l)\ .
\qe
Observe that increasing the cutoff $L$ by one adds just one additional representation. Then
using (\ref{cpn_universal_Laplacian_for_line_bundles})
we have, for the ${\bf M}_q\in {\cal F}_L\otimes {\cal F}_{L-q}^*$,
\eqa
\label{q_updown_mapping_for_cpn}
a^\dagger_{\alpha}a^\beta {\bf M}_q a^\dagger_{\beta}a^{\alpha}
&=&\left(C_2(N+1,L,L-q)-\hat{\cal L}^2\right){\bf M}_q
\nonumber\\
&&\\
a^{\alpha}a^\dagger_\beta {\bf M}_q a^{\beta}a^\dagger_{\alpha}
&=&\left(C_2(N+1,L+1,L+1-q)-\hat{\cal L}^2\right){\bf M}_q \ ,
\nonumber
\qea
where
\eq
C_2(N+1,n_{\rm L},n_{\rm R})=\frac{1}{2}\left(n_{\rm L}(n_{\rm R}+N)+n_{\rm R}(n_{\rm L}+N)\right)+\frac{N}{2(N+1)}(n_{\rm L}-n_{\rm R})^2 \ .
\qe

The operator ${a^{\beta}}^{\rm L} {a^\dagger_{\beta}}^{\rm R}$ involves the reduction of
the cutoff by one and so projects out the top representation in
(\ref{cpn_non_square_matrix_decomposition}).
The matrix is subsequently re-embedded using
${a^\dagger_{\alpha}}^{\rm L}\otimes {a^{\alpha}}^{\rm R}$.
Hence $C_2(N+1,n_{\rm L},n_{\rm R})$
is the eigenvalue of the quadratic Casimir of ${su(N+1)}$ on the representation
${(L,0,\dots,0,L-q)}$ and therefore, since increasing the cutoff $L$ by one adds just
one new representation to the
decomposition of ${\cal F}_L\otimes {\cal F}^*_{L-q}$, the
 eigenvalues of the Laplacian $\hat {\cal L}^2$ are given by $C_2(N+1,l+q,l)$
for $l=0,\dots, L-q$. More explicitly, a general matrix ${\bf M}_q$ is expanded
as in (\ref{D_polarization_expansion}), where the
${\bf D}$-polarization tensors form a basis for eigenvalues of
the Laplacian, {\it i.e.}\
\eq
\label{spectrum_for_L_L-q}
\hat {\cal L}^2 {{\bf D}_{\avec_{l+q}}}^{\bvec_{l}}
=\frac{1}{2}\left((l+q)(l+N)+l(l+q+N)
+\frac{N q^2}{N+1}\right){{\bf D}_{\avec_{l+q}}}^{\bvec_{l}}\ ,
\qe
for $l=0,\dots, n_{\rm R}$ (and $n_{\rm R}=L-q$)  and are given by
\eq
{{\bf D}_{\avec_{l+q}}}^{\bvec_{l}}=\frac{\sqrt{d_N(n_{\rm R})}}{\sqrt{Q_N(l,q,L)}}
{{{\cal P}_{\avec_{l+q},\svec_{l}}}}^{\bvec_{l},\tauvec_{l+q}}
|\tauvec_{l+q},\gvec_{n_{\rm R}-l}\rangle\langle\svec_{l},\gvec_{n_{\rm R}-l}| \ ,
\qe
which generalizes (\ref{D_polarization_tensors}) and they are normalized such that
\eq
\frac{1}{d_N(n_{\rm R})}Tr\left({({{\bf D}_{\avec_{l+q}}}^{\bvec_{l}})}^\dagger
{{\bf D}_{\svec_{l'+q'}}}^{\tauvec_{l'}}\right)
 =\delta_{ll'}\delta_{qq'}{{{\cal P}_{\bvec_{l},\svec_{l+q}}}}^{\avec_{l+q},\tauvec_{l}}
\qe
\noindent
and
\eq
Q_N(l,q,L)=\frac{l!(l+q)!}{(2l+N+q)!}\frac{(L-q-l)!(L+l+N)!}{(L-q)!L!}.
\qe
One can similarly generalize (\ref{non_square_matrix_symmetric_map_to_functions}) and
(\ref{m_q_coherentstatemap})
with
\eq
|z,n_{\rm R}\rangle\langle n_{\rm L},z|=\frac{{\cal T}^{1/2}_{L,q}(\hat{\cal L}^2,N)}{d_N(n_R)} {\bf \rho}_{L,q}({\bar z},{ z})
\qe
and now the operator ${\cal T}_{L,q}(\hat {\cal L}^2,N)$ has eigenvalues
\eq
T_{L}(l,q,N)=\frac{d_N(n_{\rm R}) R_N(l,q)}{Q_N(l,q,L)}
=\frac{L!(n_{\rm R}+N)!}{(n_{\rm R}-l)!(L+l+N)!} \ ,
\qe
where
\eq
R_N(l,q)=\frac{N!l!(l+q)!}{(2l+N+q)!}
\qe
provide the coefficients in the $\CP^N$ generalization of (\ref{Wigner-D-spin-basis}).

\section{Pseudo creation and annihilation operators}
\label{CPNstar-Fuzzy}

We would now like to obtain the generalization of the operators $\hat K_{\pm}$ for $\CP^N$.
To this end we observe that the natural generalization of
$a^{\overline{\alpha}}=a_{\alpha}=a^\beta\epsilon_{\beta\alpha}$
introduced in (\ref{A_alpha_bar_for_s2}) for $su(2)$ is obtained
by using the $\epsilon$-tensor of $su(N+1)$ contracted with $N$ oscillators. For this
to be non-zero all of the oscillators need to be distinct and hence we need to introduce
$N$ sets of oscillators $a(i)^\alpha$, with $i=1,\dots N$. The construction will then
naturally lead to dualization of the representations that occurred earlier.
To avoid this we
will use a set of oscillators $a^\imath_\alpha$ which when combined via the $\epsilon$-tensor
will give an $N$-oscillator composite operator, $A^\alpha$ with the same
transformation properties as the
$a^\alpha$ of the earlier section. By this route, the
construction leads naturally to $a^\alpha\longmapsto A^{\alpha}$
and we will have merely replaced our single oscillator by a composite one.
We begin with the composite operators
\eq
\widetilde A^{ \al}=\widetilde A_{\overline{\alpha}}
=\frac{1}{N!}\epsilon_{\overline{\alpha}\overline{\theta}_1\cdots\overline{\theta}_N}
\epsilon_{\imath_1\cdots \imath_N} a_{\theta_1}^{\imath_1}\cdots a_{\theta_N}^{\imath_N}
\qe
and
\eq \bigl(\widetilde A^{ \al}\bigr)^\dagger = \widetilde A_\al^{\dagger}
:= \frac{1}{N!}\e^{\imath_1 \cdots \imath_N}\e_{\al \ta_1 \cdots \ta_N}
(a^\dagger)_{\imath_1}^{\ta_1} \cdots (a^\dagger)_{\imath_N}^{\ta_N} \ ,
\qe
with $\alpha_i,\beta_j=1,\ldots,N+1$.
Note that $\widetilde A_{\overline{\alpha}}$ reduces to (\ref{A_alpha_bar_for_s2})
for the fuzzy sphere.

Consider the Fock space
generated freely by the subset of $N(N+1)$ oscillators
${a^\dagger}^{\alpha}_{\imath}={(a_{\alpha}^\imath)}^\dagger$, which we
denote ${\cal F}^{Total}$. These oscillators carry the anti-fundamental representation of
$u(N+1)$ and the fundamental representation of $u(N)$,
with $u(N+1)$ generators
\eq
{{\hat J}^{\alpha}}_{~~\beta}={a^\dagger}^\alpha_\imath a^\imath_\beta
\qe
and $u(N)$ generators
\eq
{\hat J_\imath}^{~\jmath}={a^\dagger}^\alpha_\imath a^\jmath_\alpha \ .
\qe
These have the common $u(1)$ generator
\eq
\hat N = {a^\dagger}^\alpha_\imath a^\imath_\alpha \ .
\qe
The generators of $u(N)$, $u(N+1)$ and $u(1)$ mutually commute, so the
Fock space ${\cal F}^{Total}$ carries a representation of
$su(N+1)\times su(N)\times u(1)$ and we can decompose it into
irreducible representations of these algebras. Fixing on an eigenvalue of
$\hat N$ fixes the total number of oscillators and we obtain the space
${\cal F}^{Total}_n$.  Due to the fact that all oscillators are
identical this space carries the symmetric representation of
$su(N(N+1))$ and when decomposed under $su(N)\times su(N+1)$ gives a
direct sum of representations. {\bf There are no branching multiplicities in this
decomposition} due to the interchangability of the identical
oscillators. Furthermore, the $su(N)$ representation is sufficient to identify
the spaces arising in the decomposition. When the total occupation number is $N L$ the
subspace ${\cal F}^{Total}_n$ contains one unique $su(n)$ singlet corresponding to $L$
copies of $\widetilde A_\al^{\dagger}$ acting on the Fock vacuum and
transforming under the $L$-fold
symmetric tensor product representation of $su(N+1)$.  The
subspace generated freely by the $\widetilde A_\al^{\dagger}$, we again denote
by ${\cal F}$ and refer to as the {\it reduced Fock space}, it
naturally decomposes into a direct sum of subspaces with fixed
eigenvalue, $L$, of the ``reduced number operator''
\eq
\hat{\cal  N}=\frac{1}{N}{(a^\dagger)}^\alpha_\imath a^\imath_\alpha=\frac{\hat N}{N}
\qe
so that
\eq
{\cal F}={\mbox{\scriptsize$\displaystyle{\boplus^{\infty}_{L=0}}$}}{\cal F}_L.
\qe
and $\hat {\cal N}$ counts the number of $\widetilde A^\dagger_\alpha$ acting on the
Fock space vacuum. These subspaces are isomorphic to those
introduced in \Ssec{\bf \ref{CPN-Fuzzy}}.

The reduced Fock space is orthogonal to the remainder so that we have
\eq
{\cal F}^{Total}= {\cal F}\oplus  {\cal F}^\perp \ .
\qe
The space ${\cal F}^{\perp}$ can further be decomposed under $su(N)$ with the leading
representation the fundamental of $su(N)$, carried by the
index $\imath$ on a single oscillator ${a^\dagger}_{\imath}^\alpha$.
In summary we can decompose ${\cal F}^{Total}$ as
\eq
\label{sum_R_Fock_space}
{\cal F}^{Total}= \oplus_{\cal R}{\cal F}_{\cal R} \ ,
\qe
where the sum is over all irreducible representations ${\cal R}$ of $u(N)$, and due to the
fact that all oscillators are identical under $u(N(N+1))$,
each representation occurs precisely once in the decomposition.

The construction described here carries over to one based on oscillators
${(a^\dagger)}^\alpha_I$ carrying representations of $u(N+1)$ and $u(k)$ with $k\leq N$.
The resulting Fock space is then decomposed as in (\ref{sum_R_Fock_space}) with the sum over
irreducible representations of $u(k)$ and again there are no multiplicities as
the symmetric representations of $u((N+1)k)$ break up into a direct sum of
representations of $u(N+1)\otimes u(k)$ without degeneracy and the
decomposition can be labeled by the representations of $u(k)$, the smaller group.

Given that we have
identified the spaces ${\cal F}_L$, then ${\cal F}\otimes {\cal F}^*$ provides the matrix algebra.
The principal distinction is
that now {\it a composite oscillator plays the role of the single oscillator}.
However, for $N \ge2$, the $\widetilde A^\dagger_\alpha$
and $\widetilde A^{\al}$ do not satisfy the Heisenberg algebra,
{\it e.g.} for $N=2$
\eqa
[\widetilde A^{ \beta},\widetilde A^\dagger_\alpha]&=&
2{\delta_\al}^\bt \left(1+\hat{\cal N}\right)
- (a^\dagger)^\bt_\imath a^\imath_\al \nonumber\\
&=& 2{\delta_\al}^\bt \left(1+\hat {\cal N}\right)
- {{\hat J}^{\beta}}_{~~\alpha}
\qea
with
\eq \hat {\cal N}:=\frac 1 2 (a^\dagger)_\imath^\delta a^\imath_\delta \ .
\qe
However, as we will see, it is straightforward to modify $\widetilde A^{\al}$ and
$\widetilde A^\dagger_\al$ to get a more suitable algebra, where they do satisfy the Heisenberg
algebra on a reduced Fock space generated by the $\widetilde A_{\alpha}^\dagger$.
We shall refer to the modified operators $A^\dagger_{\al}$ and $A^{ \al}$, defined in (\ref{composite_oscillators_defined}) and (\ref{composite_oscillators_defined2}),
as pseudo creation and pseudo annihilation operators and the subspace of the
full Fock space spanned by $\ket{\avec}$ is what we have termed the reduced Fock space.
Note: The reduced Fock space is the same as the
full Fock space only for $N=1$.

Let us examine the action of $\widetilde A^{\alpha}$ on ${\cal F}_L$.
Define
\begin{align}
\label{reduced_fockspace_states}
\ket{\widetilde\avec}
&:= \frac{1}{\sqrt{L!}}\widetilde A_{\alpha_1}^\dagger \cdots \widetilde A_{\alpha_L}^\dagger\ket 0
\nonumber\\
\quad\hbox{and}\quad
&\\
\ket{\widetilde{\hat{\avec}}_k}
&:= \frac{1}{\sqrt{(L-1)!}}\widetilde A_{\alpha_1}^\dagger \cdots \widetilde A_{\alpha_{k-1}}^\dagger \widetilde A_{\alpha_{k+1}}^\dagger \widetilde A_{\alpha_L}^\dagger\ket 0
\nonumber
\end{align}
where $\avec =(\alpha_1,\ldots,\alpha_L)$ and $\hat{\avec}_k =(\alpha_1,\ldots,\alpha_{k-1},\alpha_{k+1},\ldots,\alpha_L)$.
Then (see {appendix {\bf\ref{normalization_of_As}}}) for $L\ge 1$,
\eq \widetilde A^{ \bt}\ket{\widetilde\avec}
=\frac{c_N(L)}{\sqrt{L}}\sum_{i=1}^L\delta^{\bt}_{\al_i}\ket{\widetilde{\hat{\avec}}_i}
\qe
with
\eq c_N(L)=\frac{(N+L-1)!} {L!}.\qe
This suggests re-normalizing $\widetilde A^{ \al}$:
\eq
\label{composite_oscillators_defined}
\widetilde A^{ \al} \longrightarrow A^{ \al}
:=\widetilde A^{ \al} \frac{1}{\sqrt{c_N (\hat{\cal{N}})}}
\qe
\eq\label{composite_oscillators_defined2}
\widetilde A^\dagger_\al \longrightarrow A^\dagger_\al
:=\frac{1}{\sqrt{c_N (\hat{\cal{N}})}}
\widetilde A^\dagger_\al \ ,
\qe
so that
\eq
\label{completeness}
A^{ \bt} \ket{\avec}=\sum_{i=1}^l \dl^\bt_{\al_i} \frac{1}{\sqrt{L}}\ket{\hat{\avec}_i}
\qe
and
\eq
\label{Heisenberg_alg}
[A^{ \al} ,\, A^\dagger_{\bt}] \ket{\gvec}=\dl^\al_\bt \ket{\gvec},
\qe
where $\ket{\avec}$ is ${\ket{\widetilde\avec}}$ with $\widetilde A^\dagger_{\al_k}$ replaced with
$A^\dagger_{\al_k}$, etc.
Now (\ref{Heisenberg_alg}) implies that
$A^\dagger_{\al}$ and $A^{ \al}$ {\it act as simple creation and annihilation operators} on
${\cal F}$, the subspace of singlet representations of ${\cal F}^{Total}$.
For ${\cal F}$ we further have that
\eq
\hat {\cal N}_A{\cal F}=\hat {\cal N}{\cal F}
\qe
where $\hat {\cal N}_A=A_\al^\dagger A^\al$, so the two number operators agree on
the reduced Fock space. Of course $A^\dagger_\alpha$ and $A^\beta$ do not
satisfy a Heisenberg algebra on the whole
Fock space ${\cal F}^{Total}$.

The spaces ${\cal F}_L$ then can be identified as the space of vectors $\ket{\avec}$
satisfying the relations
\eq
\label{defn_of_pseudo_F_L}
{\hat J_\imath}^{~\jmath} \ket{\avec} = \delta_\imath^\jmath \hat {\cal N}\ket{\avec}
=L \delta_\imath^\jmath\ket{\avec}\quad {\rm and }\quad \hat {\cal N}_A\ket{\avec}=L\ket{\avec}
\qe
the first indicating the singlet nature of the state under $u(N)$.

As before, a basis for ${{\cal F}}_{n_{\rm L}}\otimes {{\cal F}}_{n_{\rm R}}^*$ is given by $\ket{\avec} \bra {\bvec}$
for all possible strings $\avec=(\alpha_1,\ldots,\alpha_{n_{\rm L}})$ and $\bvec=(\bt_1,\ldots,\bt_{n_{\rm R}})$
and the constructions of the previous section are unchanged.
As in (\ref{general_matrix_cpn}), a general matrix takes the form
\eq
\label{general_matrix_cpn_composite}
{\bf M}=\frac{1}{L!}{M^{\alpha_1\cdots\alpha_L}}_{\beta_1\cdots\beta_L}
 A_{\alpha_1}^\dagger \cdots A_{\alpha_L}^\dagger \ket 0 \bra 0 A^{\beta_1} \cdots A^{\beta_L}.
\qe
where the only distinction is that the oscillators $a^{\alpha}$ are replaced by $A^{\alpha}$.
The derivatives become
\eq
\hat{\cal L}_a={(A^\dagger\frac{\lambda_a}{2}A)}^{\rm L}-{(A^\dagger\frac{\lambda_a}{2}A)}^{\rm R}
\qe
and the Laplacian is now replaced by $\hat {\cal L}^2$ as in \Ssec{\bf \ref{CP1-Fuzzy}}
which fixes the geometry to be that of \Ssec{\bf\ref{CPN-Fuzzy}}.

Had we proceeded with a set of oscillators ${(a^\dagger)}^\imath_\alpha$,
there would be the minor consequence that ${\cal F}_L$ would transform as the $L$-fold symmetric
tensor product of the anti-fundamental rather than of the fundamental {\it i.e.}\ ${\cal F}_L$ would carry
the representation $(0,0,\dots,L)$ and hence the coefficients in the matrix algebra
${M^{\avec}}_{\bvec}$ would be replaced by ${M_{\avec}}^{\bvec}$,
so that the role of ${\bf M}$ would be replaced by that of ${\bf M}^\dagger$.

\section{Realization of the $\hat K_{\imath}$ and $\hat K_{\bar \imath}$}

The natural generalization of
(\ref{Ks_as_module_maps}) is now clear and we can define:
\eq
\begin{array}{lcllll}
\label{Kis_as_module_maps}
\hat K_{\imath}:=&{(A^\dagger_\alpha)}^{\rm L} {({({a^\dagger})}^\alpha_\imath)}^{\rm R}
&:& {\cal F}_{L}\otimes {\cal F}^*_{L}
&\longmapsto& {\cal F}_{L+1}\otimes {{\cal F}^*_{L,\imath}}\\
\hat K_{\bar\imath}:=&{(A^\alpha)}^{\rm L} {(a_\alpha^\imath)}^{\rm R}
&:&{\cal F}_{L}\otimes {\cal F}_{L}^*
&\longmapsto& {\cal F}_{L-1}\otimes {{\cal F}^*_{L,\bar\imath}} \\
\hat K_{0}:=&\frac{1}{2}(\hat{\cal N}_A^{\rm L}-\hat{\cal N}^{\rm R})&:&{\cal F}_{L}\otimes {\cal F}^*_{L}
&\longmapsto& 0 \ ,
\end{array}
\qe
where $\hat {\cal N}_A=A^\dagger_\al A^\al$ and we have denoted the subspace of Fock
space spanned by vectors of the form
\eq
{(a^\dagger)}^{\alpha}_\imath A^\dagger_{\alpha_1}\dots A^\dagger_{\alpha_L}\ket{0}
\qe
by ${\cal F}_{L,\imath}={\cal F}_L^{~\bar \imath}$. Note: Since
${(a^\dagger)}^{\alpha}_\imath A^\dagger_{\alpha}=0$ all
contractions between the $a^\imath_\alpha$ and the $A^{\alpha}$ vanish.

Observe that, quite generally,
$[\hat K_{\imath},\hat K_{\jmath}]=0=[\hat K_{\bar\imath},\hat K_{\bar\jmath}]$ and
for a generic non-square matrix ${\bf M}_q\in {\cal F}_{n_{\rm L}}\otimes{\cal F}^*_{n_{\rm R}}$ with
$q=n_{\rm L}-n_{\rm R}$, using equations (\ref{Heisenberg_alg}) and (\ref{defn_of_pseudo_F_L})
we can evaluate the commutator
\eqa
[\hat K_\imath,\,\hat K_{{\bar\jmath}} ] {\bf M}_q &=&
A^\dagger_\al A^{\bt}{\bf M}_q a^\jmath_\bt (a^\dagger)_\imath^\al
-
A^{\bt} A^\dagger_\al {\bf M}_q (a^\dagger)_\imath^\al a^\jmath_\bt \nonumber
\\&=&  A^\dagger_\al A^{ \bt}{\bf M}_q [a^\jmath_\bt,
(a^\dagger)_\imath^\al] -
{\bf M}_q {\hat J}_{\imath}^{~\jmath}
 \nonumber
\\&=& 2\dl_\imath^\jmath\hat K_0{\bf M}_q= q \delta^\jmath_\imath{\bf M}_q \ .
\qea
So we have
\eq
\label{K_i-K_j_commutator}
[\hat K_\imath,\,\hat K_{{\bar\jmath}} ] {\bf M}_q= 2\delta^{\jmath}_{\imath}\hat K_0{\bf M}_q \ .
\qe
In particular the  algebra ${\cal F}_L\otimes {\cal F}^*_L$ is
annihilated by $[\hat K_\imath,\hat K_{\bar\jmath}]$, as expected
for the action of the holonomy group on functions.

The Laplacian, $\Delta_K$, acting on ${\bf M}_q$ constructed from the
$K_\imath$ and $K_{\bar\imath}$, as demonstrated below, is naturally given by
\eq
\label{Delta_K}
\Delta_K=\frac{1}{2}\left(\hat K_{\imath}\hat K_{\bar\imath}+ \hat K_{\bar\imath}\hat K_{\imath}\right)
+\frac {2N}{N+1}\hat K_0^2 \ .
\qe
A little computation demonstrates that we can re-express $\Delta_K$ in the form
\eq
\label{Delta_K_L-J}
\Delta_K=({\hat L}_a^{\rm L}-{\hat J}_a^{\rm R})^2 \ ,
\qe
where $\hat L_a=A^\dagger\frac{\lambda_a}{2}A$ and
$\hat J_a={(a^\dagger)}_\imath\frac{{\overline\lambda}_a}{2}a^\imath$. Now it is easy to verify
that for states $\ket{\gvec}\in {\cal F}$
\eq
\hat L_a\ket{\gvec}=\hat J_a\ket{\gvec}
\qe
and hence we have for
${\bf M}_{\cal R}\in {\cal F}_L\otimes {\cal F}^*_{\cal R}$ where ${\cal R}$ is
any irreducible representation of $u(N)$,
\eq
\Delta_K{\bf M}_{\cal R}={(\hat L_a^{\rm L}-\hat J_a^{\rm R})}^2{\bf M}_{\cal R}
={(\hat J_a^{\rm L}-\hat J_a^{\rm R})}^2{\bf M}_{\cal R}\ ,
\qe
where ${(\hat J_a^{\rm L}-\hat J_a^{\rm R})}^2$ is the $su(N+1)$ quadratic Casmir and provides the
natural Laplacian on the entire space ${\cal F}^{Total}\otimes{{\cal F}^{Total}}^*$.
The eigenspaces of $\Delta_K$ are the irreducible
representations in the decomposition of ${\cal F}_L\otimes{\cal F}^*_{\cal R}$ and the eigenvalues
are those of the $su(N+1)$ quadratic Casimir in this representation.

Observe also that we can further generalize (\ref{K_i-K_j_commutator}) to obtain \eqa [\hat
K_\imath,\,\hat K_{{\bar\jmath}} ] {\bf M}_{\cal R} &=& -{\bf M}_{\cal R}{{\hat
J}_\imath}^{~\jmath}+ \delta^{\jmath}_{\imath}A^\dagger_{\alpha}A^{\alpha}{\bf M}_{\cal
R}\nonumber\\ &=&-{\bf M}_{\cal R}({{\hat J}_\imath}^{~\jmath}-\delta_\imath^\jmath\hat {\cal
N})+2\delta_\imath^\jmath\hat K_0 {\bf M}_{\cal R}\ , \qea where ${{\hat
J}_\imath}^{~\jmath}-\delta_\imath^\jmath\hat {\cal N}$ are the $su(N)$ generators and $\hat{K}_0$
is the $u(1)$ generator introduced above. Hence the effect of the commutator action $[\hat
K_\imath,\,\hat K_{{\bar\jmath}}]$ on ${\bf M}_{\cal R}$ is that of a generator of the
representation ${\cal R}$ and it effects a rotation in ${\cal R}$.

To obtain the spectrum of $\Delta_K$ by direct computation, observe that
\eqa
\hat K_\imath \ket{\gvec}\bra{\gvec}&=&0 {\rm \quad and} \\
\hat K_{{\bar\imath}}\ket{\gvec}\bra{\gvec}&=&0 \ ,\qea
which can be established by
acting on
$\ket {\bt,\gvec}=\frac{1}{\sqrt{L!}}A^\dagger_\bt A^\dagger_{\gamma_1}\cdots A^\dagger_{\gamma_L}\ket{0}$,
\eqa
\hat K_\imath \ket{\bt, \gvec}\bra{\gvec,\bt}
&=& A^\dagger_\dl A^\dagger_\bt \ket{ \gvec} \bra{\gvec}[A^{\bt} ,\,(a^\dagger)_i^\dl ]
+A^\dagger_\bt (\hat K_\imath \ket{ \gvec} \bra{\gvec})A^{\bt} \nonumber
\\ &=& A^\dagger_\bt (\hat K_\imath \ket{ \gvec} \bra{\gvec})A^{\bt}
\nonumber \qea
using antisymmetry in equation (\ref{commutator}) of  {appendix
{\bf\ref{normalization_of_As}}} and $\hat K_\imath \ket{0}\bra{0}=0$.

We can be more general and consider a non-square matrix made from a single polarization
tensor, {\it i.e.}\
${\bf \Phi}_{l_L,l_R}={f^{\avec}}_{\bvec} \ket{\avec_{l_L}, \gvec} \bra{\gvec, \bvec_{l_R}}$
where the coefficients ${f^{\avec}}_{\bvec}$ are completely traceless
and the indices $\gvec_k$, with $k=1,\ldots,n_{\rm L}-l_L=n_{\rm R}-l_R$ contracted.  Then
\eq \hat K_\imath \ket{\avec, \gvec} \bra{\gvec, \bvec} = A^\dagger_\mu A^\dagger_{\al_1} \cdots A^\dagger_{\al_{l_L}} \ket{\gvec} \bra{\gvec}
[A^{\bt_1} \cdots A^{\bt_{l_R}},\,(a^\dagger)_i^\mu ] \ .
\qe
Using (\ref{commrel}) from appendix {\bf \ref{normalization_of_As}} and
(\ref{completeness}), we now find
\eqa
\hat K_{{\bar\imath}}\hat K_\imath {\bf \Phi}_{l_L,l_R} &=&
\sum_{k=1}^{l_R} {f^{\avec}}_{\bvec} A^{\nu} A^\dagger_\mu \ket{\avec, \gvec}
\left(  \bra{\gvec, \bvec} \dl^\mu_\nu
-\bra{\gvec, \hat{\bvec}_k}A^{\mu} \dl_\nu^{\bt_k}\right) \non
\\ &=& l_R (N+n_{\rm L}+1) {\bf \Phi}_{l_L,l_R} \non \\
&& \quad- \sum_{k=1}^{l_R}{f^{\avec}}_{ \bvec} \left(\dl^{\bt_k}_\mu\ket{\avec, \gvec}+\sum_{i=1}^{n_{\rm L}-l_L}
\dl^{\bt_k}_{\g_i}\ket{\mu, \avec, \hat{\gvec}_i}  \right) \bra{\gvec, \hat{\bvec}_k, \mu } \nonumber
\\&=& l_R (l_L +N){\bf \Phi}_{l_L,l_R}.\qea
Similarly,
\eq \hat K_\imath \hat K_{{\bar\imath}}{\bf \Phi}_{l_L,l_R}=
l_L (l_R+N){\bf \Phi}_{l_L,l_R} \ . \qe
In particular for square matrices with $n_{\rm L}=n_{\rm R}=L$ and $l_l=l_R=l$,
\eq \frac 1 2 \bigl( \hat K_\imath \hat K_{\bar\imath} +
\hat K_{\bar\imath} \hat K_\imath\bigr) {{\bf Y}_{\avec_l}}^{\bvec_l} =l(l+N){{\bf Y}_{\avec_l}}^{\bvec_l},
\qe
where the polarization tensors are now built, as before, by projecting onto an irreducible representation
of $su(N+1)$, but states are built with the $A^\dagger_{\al}$.

For non-square matrices we can use the polarization tensors (again built as before)
${{\bf D}_{\avec_{l+q}}}^{\bvec_{l}}$ and with the Laplacian (\ref{Delta_K}) we have
\eq
\Delta_K{{\bf D}_{\avec_{l_{L}}}}^{\bvec_{l_{ R}}}
=
\frac{1}{2}\left(l_{ L}(l_{R}+N)+l_R(l_{ L}+N)
+\frac{N(l_{ L}-l_{ R})^2}{(N+1)}\right){{\bf D}_{\avec_{l_{ L}}}}^{\bvec_{l_{R}}} \ .
\qe
Hence putting $l_R=l$ and $q=l_L-l_R$ we agree with the spectrum (\ref{spectrum_for_L_L-q})
as derived earlier in terms of the one oscillator formulation.

Finally we can map our modules to sections of equivariant vector bundles
tensored with functions\footnote{An alternative quantization of equivariant vector
bundles using Toeplitz quantization can be found in
\cite{Hawkins:1997gj,Hawkins:1998nj,Grosse:1999ci}. } using either
${\bf \rho}_{L,q}({\bar z},{ z})$ or $|z,n_{\rm R}\rangle\langle n_{\rm L},z|$
now built from pseudo creation and
annihilation operators. These will induce the mapping
\eqa
\biggl(\frac{1}{\sqrt{\hat {\cal N}_A}}A^\dagger_\alpha\biggr)^{\rm L} \longmapsto \bar z_\alpha &\quad&
\biggl(A^{\al}\sqrt{\hat {\cal N}_A}\biggr)^{\rm L} \longmapsto \dd{}{\bar z_\al} \\
\biggl(A^\al\frac{1}{\sqrt{\hat {\cal N}_A}}\biggr)^{\rm R} \longmapsto  {z}^\al   &\quad&
 \biggl(\sqrt{\hat {\cal N}_A}A^\dagger_\al\biggr)^{\rm R}  \longmapsto\dd{}{{z}^\al} \nonumber
\label{ztoAtildedef}\qea
as expected. For elements of more general modules, such as ${\bf M}_{\cal R}$,
one needs to introduce commutative analogues of the $a_\alpha^\imath$.
For these, we will introduce $u_\alpha^\imath$ (see appendix {\bf \ref{CPN-commutative}}).
By following analogous constructions to those above, we can induce maps
\eqa
{({(a^\dagger)}^\alpha_\imath)}^{\rm L}\longmapsto  u^\alpha_\imath\quad
{\rm and }\quad {(a_\alpha^\imath)}^{\rm R}\longmapsto  \bar u_\alpha^\imath
\nonumber\\
{({a}_\alpha^\imath)}^{\rm L}\longmapsto
\frac{\partial}{\partial u^\alpha_\imath}\quad
{\rm and }\quad {({(a^\dagger)}^\alpha_\imath)}^{\rm R}\longmapsto
\frac{\partial}{\partial \bar u_\alpha^\imath}
\qea
modulo normalizations. Using appendix {\bf \ref{CPN-commutative}}
we can identify the image of the operators.

Taking the adjoint of the ${\bf M}_{\cal R}$ gives us ${}_{\overline {\cal R}}{\bf M}^\dagger
\in {\cal F}_{\overline{\cal R}}\otimes {\cal F}_L^*$. The natural operators acting on these modules
are ${}_{\imath}\hat K$ and ${}_{\bar\imath}\hat K$ which are the adjoints of
$\hat K_{\bar \imath}$ and $\hat K_{\imath}$ respectively. Explicitly they are:
\eq
{}_{\imath}\hat K= {({(a^\dagger)}^\alpha_\imath)}^{\rm L} {(A_\al^\dagger)}^{\rm R}
={(\hat K_{\bar \imath})}^\dagger\quad
{\rm and}
\quad {}_{\bar\imath}\hat K={(a^\imath_\alpha)}^{\rm L} {(A^\alpha)}^{\rm R}
={(\hat K_{\imath})}^\dagger
\qe
 One can then
see that the pairings we consider to build our Laplacians arise naturally from
action functionals of the form
\eq
\frac{Tr}{d_N({\cal R})}\left({(\hat K_{\imath}{\bf \Psi}_{\cal R})}^\dagger \hat K_\imath{\bf \Phi}_{\cal R}
+{(\hat K_{\bar \imath}{\bf \Psi}_{\cal R})}^\dagger \hat K_{\bar \imath}{\bf \Phi}_{\cal R}\right)
\qe
or
\eq
\frac{Tr}{d_N(n_{\rm L})}\left(\hat K_{\imath}{\bf \Psi}_{\cal R} {(\hat K_\imath{\bf \Phi}_{\cal R})}^\dagger
+\hat K_{\bar \imath}{\bf \Psi}_{\cal R}{(\hat K_{\bar \imath}{\bf \Phi}_{\cal R})}^\dagger\right)
\qe
where ${\bf \Psi}_{\cal R}$ and ${\bf \Phi}_{\cal R}$ are elements of the module
${\cal F}_L\otimes{\cal F}_{\cal R}^*$.
We will leave the discussion of action functionals and field theories to a separate publication.

\section{Conclusions}
\label{Conclusions}

In this paper, we have re-examined the construction of $\CP^N$. The principal novelties of the
current work are:
\begin{itemize}
\item{}The introduction of the double vacuum representation which is used
to obtain the modules representing noncommutative complex line bundles. The double vacuum,
$|0\rangle\langle 0|$ is inserted, between the creation and annihilation operators
of a normal ordered homogeneous polynomial in creation and annihilation operators.
\item{}A construction of polarization tensors that maintains equivariance at every step and
renders the polarization tensors for non-square matrices very tractable and can be
readily generalized to a large class of spaces.
\item{}We simplified the Laplacians acting on noncommutative line bundles.
\item{}We introduced a symmetric symbol density to replace the coherent state map.
This density should have an interest beyond the current work and raises questions for the future:
Does this density map positive matrices to positive functions as the
coherent state map does?
\item{}We found a new Fock space construction of $\CP^N$.
The construction allows us to access all equivariant complex vector bundles
over $\CP^N$ which are in one to one correspondence with the
representation ring of $U(N)$ (see 237 H of the Encyclopedic
Dictionary of Mathematics, second edition \cite{EDM}).
\item{}We have found composite oscillators, $A^\alpha$ of eq.~(\ref{composite_oscillators_defined}), that obey the Heisenberg algebra on the
reduced Fock space freely generated by these oscillators.
\item{}Along the way we have found a natural generalization to $su(N)$ of the
Schwinger-Jordan construction for $su(2)$ that avoids multiplicities and has
resonance with the work of Chaturvedi et al \cite{Chaturvedi:2002si,Chaturvedi:2002sj}.
\end{itemize}

The work described here opens many additional directions for investigation. The double Fock
vacuum reformulation of $\CP^N$ has already led to a simplified version of
the star product \cite{Kurkcuoglu:2006iw}. The general approach taken here leads
naturally to the construction of all flag and superflag manifolds \cite{Murray:2006pi},
where the algebra of functions is given
by ${\cal F}_{\cal R}\otimes{\cal F}^*_{\overline{\cal R}}$
with for example the representation ${\cal R}$ of $u(k)$ on the left and the
conjugate representation ${\overline{\cal R}}$ on the right realizing the algebra for
the Grassmanian $\Gr_{k;N+1}$.

Once it is appreciated that single oscillators can be replaced by composites,
many new possibilities emerge with consequences far beyond the current work.
For example, the construction should
prove useful in describing quasi-holes and quasi-particles in the higher dimensional
quantum Hall effect \cite{ZhangHu1,ZhangHu2,Karabali:2002im,Karabali:2006eg}.
Closer to home, within the framework of
noncommutative geometry, for example, one can build additional structure into the
Moyal plane by taking determinant composite oscillators, $A$, of $su(N)$, where
\eq
A={\rm det}(a)\sqrt{\frac{\hat {\cal N}!}{(\hat {\cal N}+N-1)!}}
\qe
and $a$ is the matrix of oscillators $a^\alpha_{I}$ and $\hat {\cal N}$ is the reduced
number operator
$$\hat {\cal N}=\frac{{(a^\dagger)}_{\alpha}^I a^\alpha_I}{N} \ .$$
Many further generalizations are possible. The most obvious next step is the construction of
spinor fields and their associated action functionals. We will return to this in a subsequent
article \cite{Dolan:2007}.

\bigskip

\noindent{\bf Acknowledgments} We have benefited from many discussions with our colleagues
and would  especially like to thank A.P. Balachandran, Charles Nash,
Peter Pre\v{s}najder and Christian S\"amann for their stimulating comments. The work has
been supported by Enterprise Ireland grant SC/2003/0415.

\vskip24pt
\appendix
\noindent{\Large\bf Appendices}

\section{$su(2)$ derivatives on $S^2$}
\label{commutative_s2}
For notational convenience we label the columns and rows of the entries in the matrix
differently, as $u_I^\alpha $ with $I=0,1$ and $\alpha=1,2$, for
future convenience.  Thus
\eq
\label{parametrized_su2_element}
U=\left(\begin{array}{cc}u_0^1 & u_1^1 \\ u_0^2  & u_1^2 \\
\end{array}\right) \in SU(2).
\qe
Of course not all four entries are independent,
we can write $U$ as\footnote{Indices are
raised and lowered using the unitary
metric on $\mathbb{C}^2$, $\delta^{\al\bar\bt}$ or $\delta_{\al\bar\bt}$.
Thus $\overline{(z^\al)}=(\bar z)_{\al}=\bar z_\al$.
Similarly $\overline{u_I^\alpha}=\bar u{}_{\bar I}^{\bar\al}=\bar u{}^I_\al$.}
\eq U= \left(\begin{array}{cr} z^1 &   -\bar z_2 \\ z^2 & \bar z_1 \\
\end{array}\right),\qe
where $z^\alpha$, satisfying $z^\dagger z=1$, label points on $S^3$ and
project to coordinates on $S^2$.
An alternative parameterisation is
\eq U=\left(\begin{array}{rc} \bar u{}^1_2 & u^1_1 \\
-\bar u{}^1_1  & u_1^2 \\
\end{array}\right) \qe
with $u_0^\alpha = \epsilon^{\alpha\beta} \bar u^1_\beta$, where
$\epsilon^{\alpha\beta}=-\epsilon^{\beta\alpha}$ with $\epsilon^{12}=1$.

There are three linearly independent left vector fields on $SU(2)\cong S^3$ at $U$ as well
as three linearly independent right vector fields, which
we can choose to be generated by $\frac{\sigma_a} 2 $ where $\sigma_a$ are the Pauli matrices, $a=1,2,3$,
\eq {\cal L}_a(U)=-\left(\frac {\sigma_a} 2\right) U \qquad {\cal R}_a(U)=U \left(\frac{\sigma_a} 2\right).\qe
It is straightforward to write these as differential operators
\eq -\left(\frac{\sigma_a} 2\right) U=-\frac 1 2 \left( u_I^\bt {\bigl(\sigma_a\bigr)^\alpha}_\beta
\frac{\partial}{\partial u_I^\al}\right) U, \quad
 U \left(\frac{\sigma_a} 2\right) =\frac 1 2 \left( u_I^\alpha {\bigl(\sigma_a\bigr)^I}_J
\frac{\partial}{\partial u_J^\alpha}\right) U,
\qe
where the partial derivatives are taken as though the $u_I^\al$ were independent.\footnote{In our
notation, the entries of $\sigma_a$ are necessarily labeled
differently for the right and left invariant vector fields.  This is an advantage of the notation in that
it is clear from the index structure which is left and which is right acting.}
These differential operators satisfy the $su(2)$ algebra
\eq [{\cal L}_a, {\cal L}_b]=i{\epsilon_{ab}}^c {\cal L}_c,\qquad
[{\cal R}_a,{\cal R}_b]= i{\epsilon_{ab}}^c {\cal R}_c.\qe

In the alternative raising and lowering basis (\ref{su2_in_raising_lowering_basis}) where
\hbox{$\sigma_\pm=\frac 1 2 (\sigma_1\pm i\sigma_2)$ }
and $\sigma_0=\frac 1 2 \sigma_3$,
the right invariant vector fields can be written, using $(z^1,z^2)$, as
\eqa
\label{SUtwoLeft}
{\cal L}_+ &=& -z^2\frac{\partial}{\partial z^1} + \bar z_1\frac{\partial}{\partial
\bar z_2}\nonumber \\
{\cal L}_- &=& -z^1\frac{\partial}{\partial z^2} + \bar z_2\frac{\partial}{\partial
\bar z_1}\\
{\cal L}_0 &=& -\frac 1 2 \left(z^1\frac{\partial}{\partial z^1} -
z^2\frac{\partial}{\partial z^2}\right) +
\frac 1 2 \left(\bar z_1\frac{\partial}{\partial \bar z_1} -
\bar z_2\frac{\partial}{\partial \bar z_2}\right). \nonumber
\qea

These project trivially from $S^3$ to the three linearly dependent Killing vector fields on $S^2$,
since they are invariant under a phase change $z^\alpha\rightarrow e^{i\phi}z^\alpha$.
Note that (\ref{SUtwoLeft}) decompose into two mutually commuting copies of the
$SU(2)$ algebra,
\eqa
{\cal L}_+ &=& L_+ - \overline {L_-}\nonumber \\
{\cal L}_- &=& L_- - \overline {L_+}\\
{\cal L}_0 &=& L_0 - \overline {L_0}, \nonumber
\qea
where
\eq L_+=-z^2\frac{\partial}{\partial z^1},
\qquad
L_-= -z^1\frac{\partial}{\partial z^2},\qquad
L_0= -\frac 1 2 \left(z^1\frac{\partial}{\partial z^1} - z^2\frac{\partial}{\partial z^2}\right),
\qe
associated with holomorphic and anti-holomorphic vector fields on $\CP^1$.
The right vector fields on $SU(2)$
\eqa
{\cal R}_+ &=& -z^1\frac{\partial}{\partial \bar z_2} + z^2\frac{\partial}{\partial
\bar z_1}=-\epsilon^{\alpha\beta}\,\bar u^1_\alpha \frac \partial {\partial u_1^\beta}\nonumber\\
{\cal R}_- &=& -\bar z_2\frac{\partial}{\partial z^1} +
\bar z_1\frac{\partial}{\partial z^2}=\epsilon_{\alpha\beta}u_1^\alpha \frac \partial {\partial \bar u^1_\beta}
\label{rightStwo}\\
{\cal R}_0 &=& \frac 1 2 \left(z^\alpha\frac{\partial}{\partial z^\alpha} -
\bar z_\alpha\frac{\partial}{\partial \bar z_\alpha}\right) =
\frac 1 2 \left(\bar u^1_\alpha\frac{\partial}{\partial \bar u^1_\alpha} -
u_1^\alpha\frac{\partial}{\partial u_1^\alpha}\right)\nonumber
\qea
do not project to vector fields on $S^2$, rather ${\cal R}_+$ maps functions
to $(0,1)$ tensors and ${\cal R}_-$ maps functions to $(1,0)$ tensors,
while ${\cal R}_0$ generates tangent space rotations at any point on $S^2$
that leave functions invariant, {\it i.e.}\ it generates the holonomy group $U(1)$
at the point $z^\alpha$ of $S^2$.  Nevertheless the Laplacian acting on functions
on $\CP^1$ can be written
either in terms of the left or the right vector fields,
\eqa-\nabla^2&=&\frac 1 2 \left({\cal L}_+ {\cal L}_- + {\cal L}_-{\cal L}_+\right)
+{\cal L}_0^2 =
\frac 1 2 \left({\cal R}_+ {\cal R}_- + {\cal R}_-{\cal R}_+\right)
+{\cal R}_0^2 \non \\
&=&-\partial.\bar\partial +\frac 1 2 (z.\partial)(\bar z.\bar \partial)
+\frac 1 4 (z.\partial)(z.\partial +2) + \frac 1 4 (\bar z.\bar \partial)(\bar z.\bar\partial +2) ,\qea
where $z.\partial=z^\alpha\frac \partial {\partial z^\al}$.

\section{Left and right $SU(N+1)$ invariant derivations on $\CP^N$}
\label{CPN-commutative}

Once we have mapped our modules to a representation in terms of $z^\alpha$ and $u_\imath^\alpha$
we have the following realization.

A parameterisation of an element $U$ of $G=SU(N+1)$ can be given by
\eq U=\left( \begin{array}{cccc} u_0^1 & u^1_1 & \cdots & u_N^1 \\ u_0^2 & u_1^2 & \cdots  & u_N^2 \\ \vdots & \vdots & \vdots & \vdots \\ u_0^{N+1} & u_1^{N+1} & \cdots & u_N^{N+1}\end{array}\right) \label{U} \qe
where
\eq \bar u^I_\al u_I^\bt = \delta_{\al}^\bt,\quad
\bar u^I_\al \cdot u_{J}^\al={\dl^I}_J \quad \textrm{and} \quad \e_{\al_1 \cdots \al_{N+1}} u_0^{\al_1} u_1^{\al_2} \cdots u_N^{\al_{N+1}}=1\label{paramaterization} \qe
with $I,J=0,1,\ldots N$ and $\alpha_{r}=1,\ldots,N+1$ for $r=1,\ldots,N+1$.

Acting with $G=SU(N+1)$ on $SU(N+1)$ we find that the induced left and right vector fields are given by
\eq
{\cal L}_a= -\frac 1 2 u_{I}^\bt {\bigl(\lambda_a\bigr)^\al}_\bt \dd{}{u_{I}^\al}
\quad \textrm{and} \quad
{\cal R}_a = \frac 1 2 u_{I}^\al {\bigl(\lambda_a\bigr)^{I}}_{J}\dd{}{u_{J}^\al}\ , \qe
respectively, where $\frac{\lambda_a} 2$ are the generators of $SU(N+1)$  in the
fundamental representation.  An alternative basis,
using raising and lowering operators and the Cartan subalgebra, is obtained from the completeness relation
(\ref{gell-mann_completeness})
giving
\eqa
{{\cal L}}^\al {}_{\bt}&=& -u_{I}^\al \dd{}{u_{I}^\bt} +\left(\frac {{\delta^\al}_\bt} {N+1} \right)
u_{I}^\gamma \dd{}{u_{I}^\gamma} \label{SUNleft}\\
{{\cal R}}^{I} {}_{J} &=& u_{J}^\al \dd{}{u_{I}^\al} - \left(\frac {{\delta^I}_J} {N+1} \right)
u_{K}^\al \dd{}{u_{K}^\al}.\qea
The differential operators ${{\cal L}}^\al {}_\bt$ and ${{\cal R}}^{I} {}_{J}$ separately
satisfy the commutation relations of $su(N+1)$, without recourse to (\ref{paramaterization}), and
commute with each other.

The coset space $U(N+1)/U(N)=S^{2N+1}$ can be realized by
embedding $U(N)$ in $U(N+1)$ as
\eq \left( \begin{array}{cc} e^{i \ta} & 0\\0 & h \end{array}\right)  \qe
and taking coordinates on $S^{2N+1}$ as the first column of (\ref{U}).  Restricting to functions
$x_a=\bar z \lambda_a z$ projects $S^{2N+1}$ to functions on $\CP^N$.  Special unitarity of
$U$ implies that
\eq z^\al:=u_0^\al=\e^{\al\bt_1 \cdots \bt_N} \bar u{}^1_{\bt_1} \cdots \bar u{}^N_{\bt_N}
=\frac {1} {N!} \epsilon^{\alpha\beta_1\cdots\beta_N} \epsilon_{\imath_1\cdots \imath_N}
\bar u{}^{\imath_1}_{ \bt_1} \cdots \bar u{}^{\imath_N}_{ \bt_N}\label{zdef}
\qe
and is consistent with (\ref{paramaterization}).

An equivalent description is to use
\eq \overline {z^\al}=(\bar z)_{\al}:=\bar u^0_\alpha=
\e_{\al \bt_1 \cdots \bt_N} u_1^{\bt_1} \cdots u_N^{\bt_N}
=\frac {1} {N!} \epsilon_{\alpha{\beta}_1\cdots{\beta}_N} \epsilon^{\imath_1\cdots \imath_N}
u_{\imath_1}^{\bt_1} \cdots u_{\imath_N}^{\bt_N}, \label{zbardef}
\qe
where $u_\imath^\alpha$, with $\imath=1,\ldots,N$, represent $N$ mutually
orthogonal unit vectors in ${\mathbb C}^{N+1}$
so $\bar z_{\al}$ is a hyperplane in ${\mathbb C}^{N+1}$: the set of all hyperplanes
is the Grassmanian ${\Gr}_{N;N+1}\cong {\Gr}_{1;N+1}=\CP^N$ and provides an
equivalent construction of $\CP^N$ \cite{Howe:1995md} and it is
essentially the fuzzy version of the latter that has been provided in \Ssec{\bf \ref{CPNstar-Fuzzy}}.
The $u_\imath^\alpha$ transform as a anti-fundamental representation of $SU(N)$ on the index
$\imath$ and a fundamental of $SU(N+1)$ on the index $\alpha$ while $\bar u_\al^\imath$
transform as the corresponding conjugate representations.

We would like to write the left vector fields in (\ref{SUNleft}) solely in terms of $z^\alpha$.
To this end, observe that
\eq -\left(\bar u^\imath_{\bt}  \frac\partial{\partial \bar u^\imath_{\al}}  -
\frac{{\delta^\al}_\bt}{N+1}
\bar {u}^\imath_{ \delta} \frac\partial{\partial \bar u^\imath_{ \delta}} \right) z^\gamma =
\left( z^\al \frac\partial{\partial z^\bt}
-\frac{{\delta^\al}_\bt}{N+1} z^\delta \frac\partial{\partial z^\delta}\right) z^\gamma
\qe
and the derivatives annihilate $u_\imath^\alpha$ when $\imath=1,\ldots,N$.
So we decompose $I$ into $0$ and $\imath=1,\ldots,N$ and write the left
vector fields in equation (\ref{SUNleft}) as
\eqa {{\cal L}}^\al {}_\bt &=&\left( \bar u_\bt^\imath \dd{}{\bar u_\al^\imath} -
\frac {{\delta^\al}_\bt} {N+1}  \bar u^\imath_\gamma \dd{}{\bar u_\gamma^\imath}\right) -
\left(u_{\imath}^\al \dd{}{u_{\imath}^\bt} -\frac {{\delta^\al}_\bt} {N+1}
u_{\imath}^\gamma \dd{}{u_{\imath}^\gamma}\right)\nonumber \\
&=& -\left( z^\al \frac\partial{\partial z^\bt} -\frac{{\delta^\al}_\bt}{N+1}
z^\delta \frac\partial{\partial z^\delta}\right)
+ \left( \bar z_\bt \frac\partial{\partial \bar z_\al}
-\frac{{\delta^\al}_\bt}{N+1} \bar z_\delta \frac\partial{\partial \bar z_\delta}\right)\ .
\label{CPNleft}
\qea
This last form clearly projects trivially to $\CP^N$ and, as for $\CP^1$,
decomposes into mutually commuting holomorphic and anti-holomorphic parts,
\eqa
{{\cal L}^\alpha}_\beta &=&{L^\alpha}_\beta
- \overline{{L^{\bt}}_{\al}} \ ,
\qea
where
\eq {L^\alpha}_\beta = -z^\al \frac\partial{\partial z^\bt} +\frac{{\delta^\alpha}_\beta}{N+1}
z^\delta \frac\partial{\partial z^\delta}.
\qe
The Laplacian is
\eqa -\nabla^2&=&\frac 1 2 {{\cal L}^\al}_\bt{{\cal L}^\bt}_\al\non\\
&=&-\partial.\bar\partial +\frac 1 {N+1}(z.\partial)(\bar z.\bar\partial)\\
&&\qquad+\frac N {2(N+1)}\left\{ (z.\partial)(z.\partial+N+1)
+(\bar z.\bar\partial)(\bar z.\bar\partial +N+1) \right\}.\non
\qea

For the left invariant generators of $G$ that are not in $H$, we have
\eq {\cal R}_{\bar \imath} := {\cal R}^\imath{}_0 = z^\alpha\frac\partial{\partial u_\imath^\alpha},\qquad
{\cal R}_{\imath}:={\cal R}^0{}_\imath = u_\imath^\alpha\frac\partial{\partial z^\alpha},\qe
so the $N^2$ generators of $H=U(N)$ are
\eq [{\cal R}_{\imath},{\cal R}_{\bar\jmath}]=u_\imath^\al \dd{}{u_\jmath^\al} -
{\delta^\jmath}_\imath z^\alpha\frac\partial{\partial z^\alpha}
={{\cal R}^\jmath}_\imath-\frac {{\delta^\jmath}_\imath}{N+1}
\left( N z^\alpha\frac\partial{\partial z^\al} -
u_k^\al\frac\partial{\partial u_k^\al}\right).
\qe
The commutator $[{\cal R}_\imath,{\cal R}_{\bar\jmath }]$ generates the holonomy
group of tangent space rotations, and so
vanishes on functions $f(\bar z, z)$ on $\CP^N$ as can be seen by noting that
(\ref{zdef}) gives
\eq u_\imath^\alpha \frac \partial{\partial u_\jmath^\alpha} \bar z_\beta=
{\delta_\imath}^\jmath\bar z_\beta
={\delta_\imath}^\jmath\bar z_\delta\frac{\partial}{\partial{\bar z}_\delta}\bar z_\beta\qe
so
\eq
[{\cal R}_\imath,{\cal R}_{\bar \jmath}] f(\bar z, z)=
{\delta^\jmath}_\imath \left(\bar z_\alpha\frac\partial{\partial \bar z_\alpha}-z^\alpha\frac\partial{\partial z^\alpha} \right)f(\bar z, z)=0
\qe
as required.

One can write ${\cal R}_\imath$ and ${\cal R}_{\bar\imath}$ differently
noticing that, when acting on elements of the matrix $U$,
\eq \frac{\partial}{\partial u_0^\alpha}=\frac{\partial}{\partial z^\alpha} =
\frac{\epsilon_{\alpha\beta_1\cdots\beta_N}}{N!}
\frac{\epsilon^{\imath_1\cdots\imath_N}}{N!}
\frac{\partial}{\partial \bar u^{\imath_1}_{\beta_1}}\cdots
\frac{\partial}{\partial \bar u^{\imath_N}_{\beta_N}}\ ,
\label{zdiff}
\qe
since this gives $\frac\partial {\partial z^\alpha} z^\beta ={\delta_\alpha}^\beta$
and  $\frac\partial{\partial z^\alpha} u_\imath^\beta =0$.  Thus
\eqa
{\cal R}_{\bar\imath} &=&
\e^{\al\bt_1 \cdots \bt_N} \bar{u}^1_{ \bt_1} \cdots \bar{u}^{ N}_{\bt_N} \dd{}{u_\imath^\al}
\label{KGminusH}\\
{\cal R}_{\imath}&=&
\frac 1 {N!}\e_{\al\bt_1 \cdots \bt_N} u_\imath^\al \dd{}{\bar{u}^1_{\bt_1}}
\cdots \dd{}{\bar{u}^{ N}_{\bt_N}}\nonumber
\qea
though, in contrast to the case of  $\CP^1$ in (\ref{rightStwo}), these forms are only valid for
expressions that are linear in $z^\alpha$;
a problem that will be considered in detail and rectified in the
Fock space construction of appendix {\bf \ref{normalization_of_As}}.

\section{Recovering the Heisenberg Algebra}
\label{normalization_of_As}

The annihilation and creation operators
$a^\imath_\alpha$ and $(a^\imath_\alpha)^\dagger=(a^\dagger)^\alpha_\imath$
satisfy
\eq [a^\imath_\al,(a^\dagger)_\jmath^\bt]=\delta_\jmath^\imath\delta_\al^\bt,\qe
with
$\imath=1,\ldots,N$ an $SU(N)$ index and $\alpha=1,\ldots,N+1$ an $SU(N+1)$ index.
We defined composite operators as singlets of $su(N)$ by
\eq (\widetilde A^\al)^{\dagger} =\widetilde A_\al^\dagger :=
\frac{1}{N!}\e^{\imath_1 \cdots \imath_N}\e_{\al \ta_1 \cdots \ta_N}
(a^\dagger)_{\imath_1}^{\ta_1} \cdots (a^\dagger)_{\imath_N}^{\ta_N} \ .\qe
These operators enjoy the following commutation relations:
\eq [a^\jmath_\bt,\, \tA_\al^\dagger ]
= \frac{1}{(N-1)!} \e_{\al\bt\theta_2\cdots\theta_N }\e^{\jmath \imath_1 \cdots \imath_{N-1}}
(a^\dagger)_{\imath_1}^{\ta_2} \cdots (a^\dagger)_{\imath_{N-1}}^{\ta_N},
\label{commutator}
\qe
with
\eq
\left[\hat {\cal N},\tA_\al^\dagger\right]=\tA_\al^\dagger \ ,
\qe
where
\eq \hat {\cal N}=\frac 1 N (a^\dagger)_\imath^\alpha a^\imath_\alpha\qe
is the reduced number operator and $\tA_\al$ has unit charge for the $U(1)$ associated with this
generator. Furthermore
\eq
(a^\dagger)_\imath^\gamma [a^\jmath_\gamma,\,\tA_\al^\dagger ]
= \dl^{ \jmath}_\imath \tA_\al^\dagger\ ,
\quad {\it i.e.}\quad
\left[{{\hat J}_\imath}^{~\jmath}-\delta_\imath^\jmath \hat {\cal N},\tA_\al\right]=0
\qe
reflects the fact that $\tA_\al$ is an $su(N)$ singlet and
\eq (a^\dagger)_\jmath^\gamma [a^\jmath_\bt,\,\tA_\al^\dagger ]= \dl^{ \g}_ \bt \tA_\al^\dagger
- \dl^{\g}_\al \tA_\bt^\dagger \quad {\rm or}\quad
\left[{{\hat J}^{\gamma}}_{~~\beta},\tA_\al^\dagger\right]
= \dl^{ \g}_ \bt \tA_\al^\dagger - \dl^{\g}_\al \tA_\bt^\dagger
 \label{commrel}
\qe
demonstrates that $\tA_\al$ transforms as the fundamental of ${u(N+1)}$.
Also defining the $su(N)$ singlet states $\ket{\widetilde\avec}$ and $\ket{\widetilde\avec_k}$ as in
(\ref{reduced_fockspace_states})
we therefore have
\begin{align}
{{\hat J}_\imath}^{~\jmath}\ket{\widetilde{\avec}}
&= \dl_\imath^\jmath {\hat {\cal N}}\ket{\widetilde{\avec}} \qquad {\rm and }
\nonumber\\
{\hat J}^\gamma_{~~\beta}\ket{\widetilde\avec}&=L\delta^\gamma_\beta\ket{\widetilde\avec}-\sum_{k=1}^{L}
\tA_{\beta}^\dagger\delta_{\alpha_k}^\gamma\frac{1}{\sqrt{L}}\ket{\widetilde{\hat{\avec}}_k} \ .
\label{numberrel}
\end{align}

Let us now consider the algebra of $\tA^\al$ and $\tA^\dagger_\bt$.
Define
\begin{align*}D^{\vv{\nu}}_{\vv{\mu}}(p):= D^{\nu_1 \cdots \nu_p}_{\mu_1 \cdots \mu_p}
:=\sum_{\imath_1 \cdots \imath_p  \textrm{\begin{tiny}distinct\end{tiny}}}^N
(a^\dagger)^{\nu_1}_{\imath_1}a^{\imath_1}_{\mu_1}\cdots
(a^\dagger)^{\nu_p}_{\imath_p}a^{\imath_p}_{\mu_p}~.
\end{align*}
First observe the lower and upper extreme cases:
$$ D^{\nu}_\mu(1)={{(a^\dagger)}}_\imath^\nu a^\imath_{\mu} ={{\hat J}^\nu}_{~~\mu} \ ,$$
are the generators of $U(N+1)$ and
$$ \dl^{\al \vv{\mu}}_{\bt \vv{\nu}} D^{\vv{\nu}}_{\vv{\mu}}(N)=
N! \tA^\dagger_\bt \tA^\al \ ,$$
where $\dl^{\al \vv{\mu}}_{\bt \vv{\nu}}$ is the $p+1$-delta symbol,
the $p$-delta symbol being defined via
\eq
\e^{\al_1 \cdots \al_{p}\g_{p+1}\cdots \g_{N+1}} \e_{\bt_1 \cdots \bt_{p}\g_{p+1}\cdots \g_{N+1}}
=(N+1-p)!\dl^{\al_1 \cdots \al_{p}}_{\bt_1 \cdots \bt_{p}}
\qe
and explicitly
\eq
\qquad \dl^{\al_1 \cdots \al_{p}}_{\bt_1 \cdots \bt_{p}}
=\delta^{\al_1}_{\bt_1}\delta^{\al_2}_{\bt_2}\dots\delta^{\al_p}_{\bt_p}
-\delta^{\al_1}_{\bt_2}\delta^{\al_2}_{\bt_1}\dots\delta^{\al_p}_{\bt_p}+\dots
\qe
with $p!$ terms.

Now separating ${\hat J^{\nu_{p+1}}}_{~~\mu_{p+1}}
={(a^\dagger)}^{\nu_{p+1}}_\imath a^\imath_{\nu_{p+1}}$ as
\eq
\hat J^{\nu_{p+1}}_{~~\mu_{p+1}}= (a^\dagger)^{\nu_{p+1}}_{\imath_1} a^{\imath_1}_{\mu_{p+1}}
+ \cdots (a^\dagger)^{\nu_{p+1}}_{\imath_p} a^{\imath_p}_{\mu_{p+1}}
+\sum_{\imath_{p+1}\notin \{\imath_1,\ldots,\imath_p \}}
(a^\dagger)^{\nu_{p+1}}_{\imath_{p+1}} a^{\imath_{p+1}}_{\mu_{p+1}}
\qe
and substituting into
\eq
\dl^{\alpha\mu_1 \cdots \mu_{p}\mu_{p+1}}_{\beta\nu_1 \cdots \nu_{p}\nu_{p+1}}
D^{\nu_1 \cdots \nu_p}_{\mu_1 \cdots \mu_p}{\hat J^{\nu_{p+1}}}_{~~\mu_{p+1}}
\qe
we see that due to antisymmetry of the $p$-delta symbol we have
\eq
\label{deltapDJ}
\dl^{\alpha\mu_1 \cdots \mu_{p}\mu_{p+1}}_{\beta\nu_1 \cdots \nu_{p}\nu_{p+1}}
(a^\dagger)^{\nu_1}_{\imath_1}a^{\imath_1}_{\mu_1}\cdots
(a^\dagger)^{\nu_p}_{\imath_p}a^{\imath_p}_{\mu_j}
\left({{(a^\dagger)}^{\nu_{p+1}}_{\imath_k}}a^{\imath_k}_{\mu_{p+1}}
+\delta^{\nu_{p+1}}_{\mu_{p+1}}\right)=0
\qe
for $k=1,2,...,p$ and so we obtain the recursion relation
\begin{align}  \dl^{\al \vv{\mu}}_{\bt \vv{\nu}} D^{\vv{\nu}}_{\vv{\mu}}(p+1)
=\dl^{\al \vv{\mu}\mu_{p+1}}_{\bt \vv{\nu}\nu_{p+1}}
D^{\vv{\nu}}_{\vv{\mu}}(p) \left( p \dl^{\nu_{p+1}}_{\mu_{p+1}}+\hat J^{\nu_{p+1}}_{~~\mu_{p+1}}\right) \nonumber
\end{align}
which gives
\eq
\label{tAdagtA}
\tA^\dagger_{\bt} \tA^{\al}=
\frac{1} {N!} \dl^{\al \vv{\mu}}_{\bt \vv{\nu}}\prod_{p=1}^{N}
\left( (p-1)\delta_{\mu_p}^{\nu_p}+\hat J_{~~\mu_p}^{\nu_p}\right)\ .\qe

Similarly defining $\underline{D}^{\vv{\nu}}_{\vv{\mu}}$ by interchanging the roles of
creation and annihilation operators in $D^{\vv{\nu}}_{\vv{\mu}}$, one obtains the recursion
relation
\eq
\label{test}
\dl^{\al \vv{\mu}}_{\bt \vv{\nu}} {\underline{D}}^{\vv{\nu}}_{\vv{\mu}}(p+1)
=\dl^{\al \vv{\mu}\mu_{p+1}}_{\bt \vv{\nu}\nu_{p+1}}
{\underline{D}}^{\vv{\nu}}_{\vv{\mu}}(p)
{\left((N-p)\dl^{\nu_{p+1}}_{\mu_{p+1}}+\hat J^{\nu_{p+1}}_{~~\mu_{p+1}}\right)}
\ .
\qe
With the initial condition
\eq
\underline{D}^\nu_\mu(1)={\hat J^{\nu}}_{~~\mu}+N\delta^\nu_\mu \ ,
\qe
itterating (\ref{test}) yields
\eq
\label{tAtAdag}
\tA^{\al}\tA^\dagger_{\bt}=\frac{1} {N!}\dl^{\al \vv{\mu}}_{\bt \vv{\nu}} \prod_{p=1}^{N}
\left( p\delta_{\mu_p}^{\nu_p}+\hat J_{~~\mu_p}^{\nu_p}\right)\ .
\qe
Thus
we have that $\tA^\dagger_{\al}\tA^\al$ and
$\tA^\al\tA^\dagger_{\al}$ can be expressed as polynomials in the Casimirs of $u(N+1)$ up to $C_{N}$.

Furthermore, the commutator is
\eq
\label{AAdag_commutator}
[\tA^{\al},\,\tA^\dagger_{\bt}]=
\frac{1}{(N-1)!}\dl^{\al \vv{\mu}}_{\bt \vv{\nu}}
\prod_{p=1}^{N-1} \left( p\delta_{\mu_p}^{\nu_p}+\hat J_{~~\mu_p}^{\nu_p}\right)\dl^{\nu_N}_{\mu_N}~,
\nonumber
\qe
and the commutator is a polynomial in the Casimirs, $C_i$, of $u(N+1)$ up to $C_{N-1}$.

We can use this result to determine how $\tA^\al$ behaves on the reduced Fock space $\ket{\gvec}$.
We contract all
$\tA^\dagger_\al$ with an arbitrary vector $x^\al$ to obtain $X=x^\al\tA^\dagger_\al$. Then,
\eq
\label{Atilde_onX_N}
\tA^\al \ket{X^{L}}=\frac{x^\bt}{N!\sqrt{L}}\dl^{\al \vv{\mu}}_{\bt \vv{\nu}}
\prod_{p=1}^{N}\left( p\delta_{\mu_p}^{\nu_p}+\hat J_{~~\mu_p}^{\nu_p}\right)\ket{X^{L-1}} \ . \qe
Next, we use (\ref{commrel}) to see that
\eq
\label{commutator_X}
[\hat J^\nu_{~~\mu}-\dl^\nu_{~\mu}\hat {\cal N}, X]=-x^\nu \tA^\dagger_\mu \ ,
\qe
which allows us to substitute for $\hat J^\nu_{~~\mu}$ in (\ref{Atilde_onX_N}); then observing that
the contribution from the right hand side of (\ref{commutator_X}) gives zero,
due to antisymmetrization of $x^\mu$ with $x^\beta$, we have that ${\hat J^{\nu}}_{~~\mu}$
is replaced with $\hat {\cal N}\delta^\nu_\mu$ when acting on $\ket{X^{L-1}}$ in (\ref{Atilde_onX_N}) so that
\begin{align}
\tA^\al \ket{X^{L}}=\,\frac{x^\bt}{N!\sqrt{L}}\dl^{\al \vv{\mu}}_{\bt \vv{\nu}}
\prod_{p=1}^{N}\left( (p+\hat {\cal N})\delta_{\mu_p}^{\nu_p}\right)\ket{X^{L-1}}
=\,x^\al \frac{(\hat {\cal N}+N)!}{\hat {\cal N}!\sqrt{L}}\ket{X^{L-1}}\ ~.\end{align}
Taking $L$ derivatives with respect to $x^\g$ we find
\eq \tA^\al \ket{\gvec}=\frac{(L+N-1)!}{L!\sqrt{L}}\sum_{i=1}^{L}\dl^\al_{~\g_i}\ket{\hat{\gvec}_i}. \qe

Now defining
\eq
A^{\al}:=\tA^\al\sqrt{\frac{\hat {\cal N}!}{(\hat {\cal N}+N-1)!}}
\qe
we have, when acting on states $\ket{\gvec}$ of the reduced Fock space,
\eq
A^\dagger_\al A^{\al}\ket{\gvec}=\hat {\cal N}\ket{\gvec}
\qe
and
\eq
[A^{\al},A^\dagger_\bt]\ket{\gvec}=\delta^{\al}_\bt\ket{\gvec} \ .
\qe
Hence we find that $A^\al$ and $A^\dagger_\beta$ obey the Heisenberg Algebra.

Form the above we can write
\eq
A^\dagger_\bt A^{\al}=\hat {\cal N}\frac{1}{N!}\dl^{\al\vv{\mu}}_{\bt \vv{\nu}}
\prod_{p=0}^{N-1} {\left(\frac{ p+\hat J}{p+\hat {\cal N}}\right)\kern-4pt}_{\mu_p}^{\nu_p}
\qe
and
\eq
A^\al A^\dagger_{\beta}={(\hat {\cal N}+1)}\frac{1}{N!}\dl^{\al\vv{\mu}}_{\bt \vv{\nu}}
\prod_{p=1}^{N} {\left(\frac{ p+\hat J}{p+\hat {\cal N}}\right)\kern-4pt}_{\mu_p}^{\nu_p}
\qe
so that
\eq
[A^{\al},A^\dagger_\bt]=\frac{1}{N!}\dl^{\al \vv{\mu}}_{\bt \vv{\nu}}
\prod_{p=1}^{N-1} {\left(\frac{ p+\hat J}{p+\hat {\cal N}}\right)\kern-4pt}_{\mu_p}^{\nu_p}
{\left(\frac{N(\hat {\cal N}+1)+(1-N)\hat J}{N+\hat {\cal N}}\right)\kern-4pt}^{\nu_N}_{\mu_N}~.
\qe
Also observe that (\ref{numberrel}) becomes
\eq
({\hat J}^\gamma_{~~\beta}+A_{\beta}^\dagger A^{\gamma})\ket{\avec}
=\hat{\cal N}\delta^\gamma_\beta\ket{\avec} \ .
\qe

\comment{

}

\bibliographystyle{JHEP}
\bibliography{bibfile}

\providecommand{\href}[2]{#2}\begingroup\raggedright\begin{thebibliography}{10}

\bibitem{Connes:1994yd}
A.~Connes, {\em Noncommutative Geometry}.
\newblock Academic Press, 1994.

\bibitem{GraciaBondia:2001tr}
J.~M. Gracia-Bondia, J.~C. Varilly, and H.~Figueroa, {\em Elements of
  noncommutative geometry}.
\newblock Birkh{\"a}user, 2001.

\bibitem{Madore:2000aq}
J.~Madore, {\it An introduction to noncommutative differential geometry and
  physical applications},  {\em Lond. Math. Soc. Lect. Note Ser.} {\bf 257}
  (2000) 1--371.

\bibitem{Landi:1997sh}
G.~Landi, {\em An introduction to noncommutative spaces and their geometry},
  vol.~51 of {\em Lecture Notes in Physics monographs}.
\newblock Springer Berlin/Heidelberg, 2002.
\newblock [hep-th/9701078].

\bibitem{Balachandran:2005eb}
A.~P. Balachandran, G.~Mangano, A.~Pinzul, and S.~Vaidya, {\it Spin and
  statistics on the {G}roenwald-{M}oyal plane: Pauli-forbidden levels and
  transitions},  {\em Int. J. Mod. Phys.} {\bf A21} (2006) 3111--3126,
  [\href{http://xxx.lanl.gov/abs/hep-th/0508002}{{\tt hep-th/0508002}}].

\bibitem{Doplicher:1994tu}
S.~Doplicher, K.~Fredenhagen, and J.~E. Roberts, {\it The quantum structure of
  space-time at the {P}lanck scale and quantum fields},  {\em Commun. Math.
  Phys.} {\bf 172} (1995) 187--220,
  [\href{http://xxx.lanl.gov/abs/hep-th/0303037}{{\tt hep-th/0303037}}].

\bibitem{Seiberg:1999vs}
N.~Seiberg and E.~Witten, {\it String theory and noncommutative geometry},
  {\em JHEP} {\bf 09} (1999) 032,
  [\href{http://xxx.lanl.gov/abs/hep-th/9908142}{{\tt hep-th/9908142}}].

\bibitem{Aschieri:2003vy}
P.~Aschieri, J.~Madore, P.~Manousselis, and G.~Zoupanos, {\it Dimensional
  reduction over fuzzy coset spaces},  {\em JHEP} {\bf 04} (2004) 034,
  [\href{http://xxx.lanl.gov/abs/hep-th/0310072}{{\tt hep-th/0310072}}].

\bibitem{Dolan:2002af}
B.~P. Dolan and C.~Nash, {\it Chiral fermions and spin(c) structures on matrix
  approximations to manifolds},  {\em JHEP} {\bf 07} (2002) 057,
  [\href{http://xxx.lanl.gov/abs/hep-th/0207007}{{\tt hep-th/0207007}}].

\bibitem{Dolan:2002ck}
B.~P. Dolan and C.~Nash, {\it The standard model fermion spectrum from complex
  projective spaces},  {\em JHEP} {\bf 10} (2002) 041,
  [\href{http://xxx.lanl.gov/abs/hep-th/0207078}{{\tt hep-th/0207078}}].

\bibitem{Balachandran:2005ew}
A.~P. Balachandran, S.~Kurkcuoglu, and S.~Vaidya, {\it Lectures on fuzzy and
  fuzzy {SUSY} physics},  \href{http://xxx.lanl.gov/abs/hep-th/0511114}{{\tt
  hep-th/0511114}}.

\bibitem{Douglas:2001ba}
M.~R. Douglas and N.~A. Nekrasov, {\it Noncommutative field theory},  {\em Rev.
  Mod. Phys.} {\bf 73} (2001) 977--1029,
  [\href{http://xxx.lanl.gov/abs/hep-th/0106048}{{\tt hep-th/0106048}}].

\bibitem{Szabo:2001kg}
R.~J. Szabo, {\it Quantum field theory on noncommutative spaces},  {\em Phys.
  Rept.} {\bf 378} (2003) 207--299,
  [\href{http://xxx.lanl.gov/abs/hep-th/0109162}{{\tt hep-th/0109162}}].

\bibitem{Taylor:2001vb}
W.~Taylor, {\it M(atrix) theory: Matrix quantum mechanics as a fundamental
  theory},  {\em Rev. Mod. Phys.} {\bf 73} (2001) 419--462,
  [\href{http://xxx.lanl.gov/abs/hep-th/0101126}{{\tt hep-th/0101126}}].

\bibitem{Karabali:2006eg}
D.~Karabali and V.~P. Nair, {\it Quantum {H}all effect in higher dimensions,
  matrix models and fuzzy geometry},  {\em J. Phys.} {\bf A39} (2006)
  12735--12764, [\href{http://xxx.lanl.gov/abs/hep-th/0606161}{{\tt
  hep-th/0606161}}].

\bibitem{Balachandran:1999qu}
A.~P. Balachandran, T.~R. Govindarajan, and B.~Ydri, {\it The fermion doubling
  problem and noncommutative geometry},  {\em Mod. Phys. Lett.} {\bf A15}
  (2000) 1279, [\href{http://xxx.lanl.gov/abs/hep-th/9911087}{{\tt
  hep-th/9911087}}].

\bibitem{Martin:2004un}
X.~Martin, {\it A matrix phase for the $\phi^4$ scalar field on the fuzzy
  sphere},  {\em JHEP} {\bf 04} (2004) 077,
  [\href{http://xxx.lanl.gov/abs/hep-th/0402230}{{\tt hep-th/0402230}}].

\bibitem{GarciaFlores:2005xc}
F.~Garcia~Flores, D.~O'Connor, and X.~Martin, {\it Simulating the scalar field
  on the fuzzy sphere},  {\em PoS} {\bf LAT2005} (2006) 262,
  [\href{http://xxx.lanl.gov/abs/hep-lat/0601012}{{\tt hep-lat/0601012}}].

\bibitem{Medina:2005su}
J.~Medina, W.~Bietenholz, F.~Hofheinz, and D.~O'Connor, {\it Field theory
  simulations on a fuzzy sphere: An alternative to the lattice},  {\em PoS}
  {\bf LAT2005} (2006) 263,
  [\href{http://xxx.lanl.gov/abs/hep-lat/0509162}{{\tt hep-lat/0509162}}].

\bibitem{Panero:2006bx}
M.~Panero, {\it Numerical simulations of a non-commutative theory: The scalar
  model on the fuzzy sphere},  {\em JHEP} {\bf 05} (2007) 082,
  [\href{http://xxx.lanl.gov/abs/hep-th/0608202}{{\tt hep-th/0608202}}].

\bibitem{Panero:2006cs}
M.~Panero, {\it Quantum field theory in a non-commutative space: Theoretical
  predictions and numerical results on the fuzzy sphere},  {\em SIGMA} {\bf 2}
  (2006) 081, [\href{http://xxx.lanl.gov/abs/hep-th/0609205}{{\tt
  hep-th/0609205}}].

\bibitem{Grosse:2001ss}
H.~Grosse, M.~Maceda, J.~Madore, and H.~Steinacker, {\it Fuzzy instantons},
  {\em Int. J. Mod. Phys.} {\bf A17} (2002) 2095,
  [\href{http://xxx.lanl.gov/abs/hep-th/0107068}{{\tt hep-th/0107068}}].

\bibitem{Steinacker:2003sd}
H.~Steinacker, {\it Quantized gauge theory on the fuzzy sphere as random matrix
  model},  {\em Nucl. Phys.} {\bf B679} (2004) 66--98,
  [\href{http://xxx.lanl.gov/abs/hep-th/0307075}{{\tt hep-th/0307075}}].

\bibitem{Castro-Villarreal:2004vh}
P.~Castro-Villarreal, R.~Delgadillo-Blando, and B.~Ydri, {\it A gauge-invariant
  {UV-IR} mixing and the corresponding phase transition for {U(1)} fields on
  the fuzzy sphere},  {\em Nucl. Phys.} {\bf B704} (2005) 111--153,
  [\href{http://xxx.lanl.gov/abs/hep-th/0405201}{{\tt hep-th/0405201}}].

\bibitem{Grosse:2004wm}
H.~Grosse and H.~Steinacker, {\it Finite gauge theory on fuzzy ${CP}^2$},  {\em
  Nucl. Phys.} {\bf B707} (2005) 145--198,
  [\href{http://xxx.lanl.gov/abs/hep-th/0407089}{{\tt hep-th/0407089}}].

\bibitem{O'Connor:2006wv}
D.~O'Connor and B.~Ydri, {\it Monte {C}arlo simulation of a {NC} gauge theory
  on the fuzzy sphere},  {\em JHEP} {\bf 11} (2006) 016,
  [\href{http://xxx.lanl.gov/abs/hep-lat/0606013}{{\tt hep-lat/0606013}}].

\bibitem{Bietenholz:2006cz}
W.~Bietenholz, J.~Nishimura, Y.~Susaki, and J.~Volkholz, {\it A
  non-perturbative study of 4d {U(1)} non-commutative gauge theory: The fate of
  one-loop instability},  {\em JHEP} {\bf 10} (2006) 042,
  [\href{http://xxx.lanl.gov/abs/hep-th/0608072}{{\tt hep-th/0608072}}].

\bibitem{Berezin:1974du}
F.~A. Berezin, {\it General concept of quantization},  {\em Commun. Math.
  Phys.} {\bf 40} (1975) 153--174.

\bibitem{Hoppe:1982}
J.~Hoppe, {\em Quantum Theory of a Massless Relativistic Surface and a Two
  Dimensional Bound State Problem}.
\newblock PhD thesis, MIT, 1982.

\bibitem{Madore:1991bw}
J.~Madore, {\it The fuzzy sphere},  {\em Class. Quant. Grav.} {\bf 9} (1992)
  69--88.

\bibitem{Arnlind:2006ux}
J.~Arnlind, M.~Bordemann, L.~Hofer, J.~Hoppe, and H.~Shimada, {\it Fuzzy
  {R}iemann surfaces},  \href{http://xxx.lanl.gov/abs/hep-th/0602290}{{\tt
  hep-th/0602290}}.

\bibitem{Balachandran:2001dd}
A.~P. Balachandran, B.~P. Dolan, J.-H. Lee, X.~Martin, and D.~O'Connor, {\it
  Fuzzy complex projective spaces and their star-products},  {\em J. Geom.
  Phys.} {\bf 43} (2002) 184--204,
  [\href{http://xxx.lanl.gov/abs/hep-th/0107099}{{\tt hep-th/0107099}}].

\bibitem{Dolan:2001mi}
B.~P. Dolan and O.~Jahn, {\it Fuzzy complex {G}rassmannian spaces and their
  star products},  {\em Int. J. Mod. Phys.} {\bf A18} (2003) 1935--1958,
  [\href{http://xxx.lanl.gov/abs/hep-th/0111020}{{\tt hep-th/0111020}}].

\bibitem{Dolan:2003th}
B.~P. Dolan, D.~O'Connor, and P.~Pre$\check{\mathrm{s}}$najder, {\it Fuzzy
  complex quadrics and spheres},  {\em JHEP} {\bf 02} (2004) 055,
  [\href{http://xxx.lanl.gov/abs/hep-th/0312190}{{\tt hep-th/0312190}}].

\bibitem{Ramgoolam:2001zx}
S.~Ramgoolam, {\it On spherical harmonics for fuzzy spheres in diverse
  dimensions},  {\em Nucl. Phys.} {\bf B610} (2001) 461--488,
  [\href{http://xxx.lanl.gov/abs/hep-th/0105006}{{\tt hep-th/0105006}}].

\bibitem{Dolan:2003kq}
B.~P. Dolan and D.~O'Connor, {\it A fuzzy three sphere and fuzzy tori},  {\em
  JHEP} {\bf 10} (2003) 060,
  [\href{http://xxx.lanl.gov/abs/hep-th/0306231}{{\tt hep-th/0306231}}].

\bibitem{Grosse:1995jt}
H.~Grosse, C.~Klimcik, and P.~Pre$\check{\mathrm{s}}$najder, {\it Topologically
  nontrivial field configurations in noncommutative geometry},  {\em Commun.
  Math. Phys.} {\bf 178} (1996) 507--526,
  [\href{http://xxx.lanl.gov/abs/hep-th/9510083}{{\tt hep-th/9510083}}].

\bibitem{Presnajder:1999ky}
P.~Pre$\check{\mathrm{s}}$najder, {\it The origin of chiral anomaly and the
  noncommutative geometry},  {\em J. Math. Phys.} {\bf 41} (2000) 2789--2804,
  [\href{http://xxx.lanl.gov/abs/hep-th/9912050}{{\tt hep-th/9912050}}].

\bibitem{Perelomov:1986tf}
A.~M. Perelomov, {\em Generalized coherent states and their applications}.
\newblock Springer Berlin, 1986.

\bibitem{Berezin:1975:Izv}
F.~Berezin, {\it Quantization on complex symmetric spaces},  {\em Izvestija}
  {\bf 9} (1975) 341.

\bibitem{Dolan:2001gn}
B.~P. Dolan, D.~O'Connor, and P.~Pre$\check{\mathrm{s}}$najder, {\it Matrix
  $\phi^4$ models on the fuzzy sphere and their continuum limits},  {\em JHEP}
  {\bf 03} (2002) 013, [\href{http://xxx.lanl.gov/abs/hep-th/0109084}{{\tt
  hep-th/0109084}}].

\bibitem{Kurkcuoglu:2006iw}
S.~Kurkcuoglu and C.~Saemann, {\it Drinfeld twist and general relativity with
  fuzzy spaces},  {\em Class. Quant. Grav.} {\bf 24} (2007) 291--312,
  [\href{http://xxx.lanl.gov/abs/hep-th/0606197}{{\tt hep-th/0606197}}].

\bibitem{Baez:1998he}
S.~Baez, A.~P. Balachandran, B.~Ydri, and S.~Vaidya, {\it Monopoles and
  solitons in fuzzy physics},  {\em Commun. Math. Phys.} {\bf 208} (2000)
  787--798, [\href{http://xxx.lanl.gov/abs/hep-th/9811169}{{\tt
  hep-th/9811169}}].

\bibitem{Balachandran:1999hx}
A.~P. Balachandran and S.~Vaidya, {\it Instantons and chiral anomaly in fuzzy
  physics},  {\em Int. J. Mod. Phys.} {\bf A16} (2001) 17--40,
  [\href{http://xxx.lanl.gov/abs/hep-th/9910129}{{\tt hep-th/9910129}}].

\bibitem{Grosse:2001qt}
H.~Grosse, C.~W. Rupp, and A.~Strohmaier, {\it Fuzzy line bundles, the {C}hern
  character and topological charges over the fuzzy sphere},  {\em J. Geom.
  Phys.} {\bf 42} (2002) 54--63,
  [\href{http://xxx.lanl.gov/abs/math-ph/0105033}{{\tt math-ph/0105033}}].

\bibitem{CarowWatamura:2004ct}
U.~Carow-Watamura, H.~Steinacker, and S.~Watamura, {\it Monopole bundles over
  fuzzy complex projective spaces},  {\em J. Geom. Phys.} {\bf 54} (2005)
  373--399, [\href{http://xxx.lanl.gov/abs/hep-th/0404130}{{\tt
  hep-th/0404130}}].

\bibitem{Hawkins:1997gj}
E.~Hawkins, {\it Quantization of equivariant vector bundles},  {\em Commun.
  Math. Phys.} {\bf 202} (1999) 517--546,
  [\href{http://xxx.lanl.gov/abs/q-alg/9708030}{{\tt q-alg/9708030}}].

\bibitem{Hawkins:1998nj}
E.~Hawkins, {\it Geometric quantization of vector bundles},  {\em Commun. Math.
  Phys.} {\bf 215} (2000) 409--432,
  [\href{http://xxx.lanl.gov/abs/math/9808116}{{\tt math/9808116}}].

\bibitem{Grosse:1999ci}
H.~Grosse and A.~Strohmaier, {\it Noncommutative geometry and the
  regularization problem of 4d quantum field theory},  {\em Lett. Math. Phys.}
  {\bf 48} (1999) 163--179, [\href{http://xxx.lanl.gov/abs/hep-th/9902138}{{\tt
  hep-th/9902138}}].

\bibitem{EDM}
{Mathematical Society of Japan}, {\em Encyclopedic Dictionary of Mathematics}.
\newblock MIT Press, second~ed., 1987.

\bibitem{Chaturvedi:2002si}
S.~Chaturvedi and N.~Mukunda, {\it The {S}chwinger {SU(3)} construction - {I}:
  Multiplicity problem and relation to induced representations},  {\em J. Math.
  Phys.} {\bf 43} (2002) 5262--5277,
  [\href{http://xxx.lanl.gov/abs/quant-ph/0204119}{{\tt quant-ph/0204119}}].

\bibitem{Chaturvedi:2002sj}
S.~Chaturvedi and N.~Mukunda, {\it The {S}chwinger {SU(3)} construction - {II}:
  Relations between {H}eisenberg-{W}eyl and {SU(3)} coherent states},  {\em J.
  Math. Phys.} {\bf 43} (2002) 5278--5309,
  [\href{http://xxx.lanl.gov/abs/quant-ph/0204120}{{\tt quant-ph/0204120}}].

\bibitem{Murray:2006pi}
S.~Murray and C.~Saemann, {\it Quantization of flag manifolds and their
  supersymmetric extensions},  {\em ATMP} {\bf 12} (2008), no.~3
  [\href{http://xxx.lanl.gov/abs/hep-th/0611328}{{\tt hep-th/0611328}}].

\bibitem{ZhangHu1}
J.~Hu and S.-C. Zhang, {\it A four dimensional generalization of the quantum
  {H}all effect},  {\em Science} {\bf 294} (2001) 823,
  [\href{http://xxx.lanl.gov/abs/cond-mat/0110572}{{\tt cond-mat/0110572}}].

\bibitem{ZhangHu2}
J.~Hu and S.-C. Zhang, {\it Collective excitations at the boundary of a quantum
  {H}all droplet},  {\em Phys. Rev.} {\bf B66} (2002) 125301,
  [\href{http://xxx.lanl.gov/abs/cond-mat/0112432}{{\tt cond-mat/0112432}}].

\bibitem{Karabali:2002im}
D.~Karabali and V.~P. Nair, {\it Quantum hall effect in higher dimensions},
  {\em Nucl. Phys.} {\bf B641} (2002) 533--546,
  [\href{http://xxx.lanl.gov/abs/hep-th/0203264}{{\tt hep-th/0203264}}].

\bibitem{Dolan:2007}
B.~P. Dolan, I.~Huet, S.~Murray, and D.~O'Connor, {\it A universal {D}irac
  operator and noncommutative spin bundles over fuzzy complex projective
  spaces},  {\em JHEP} {\bf 03} (2008) 029,
  [\href{http://xxx.lanl.gov/abs/arXiv{:}0711.1347 [hep-th]}{{\tt
  arXiv{:}0711.1347 [hep-th]}}].

\bibitem{Howe:1995md}
P.~S. Howe and G.~G. Hartwell, {\it A superspace survey},  {\em Class. Quant.
  Grav.} {\bf 12} (1995) 1823--1880.

\end{thebibliography}\endgroup

\end{document}